\documentclass[format=acmsmall, review=false, screen=true]{acmart}

\usepackage{booktabs} 

\usepackage[ruled]{algorithm2e} 

\SetAlFnt{\small}
\SetAlCapFnt{\small}
\SetAlCapNameFnt{\small}
\SetAlCapHSkip{0pt}
\IncMargin{-\parindent}
\usepackage{enumitem}
\usepackage{textcomp}
\RequirePackage{tcolorbox}
\usepackage{graphicx}  
\usepackage{balance}       
\usepackage{multirow}
\usepackage{rotating}
\usepackage{subcaption}
\usepackage{refcount}
\usepackage{paralist}
\usepackage{colortbl}
\usepackage{threeparttable}
\usepackage{tikz}
\usetikzlibrary{arrows, fit,positioning}
\usepackage{pgfplots}
\pgfplotsset{width=7cm,compat=1.8}

 \usepackage[]{footmisc}

\usepackage{array} 
\graphicspath{{images/}}
\newcommand\TODO[1]{\textcolor{red}{#1}}

\usepackage[font=small,labelfont=bf]{caption}

\makeatother

\setcopyright{acmcopyright}
\acmJournal{PACMHCI}
\acmYear{2018} \acmVolume{2} \acmNumber{CSCW} \acmArticle{117} \acmMonth{1} \acmPrice{15.00}\acmDOI{10.1145/3274386}

\begin{document}
\title[Incivility]{Opinion Conflicts: An Effective Route to Detect Incivility in Twitter}

\author{Suman Kalyan Maity}
\affiliation{
  \institution{Northwestern University}
  \department{Kellogg School of Management and Northwestern Institute on Complex Systems}
  \city{Evanston}
  \state{IL}
  \postcode{60208}
  \country{USA}}
\email{suman.maity@kellogg.northwestern.edu}
\author{Aishik Chakraborty}
\affiliation{
  \institution{McGill University}
  \department{School of Computer Science}
  \city{Montreal}
  \country{Canada}
}
\email{chakraborty.aishik@gmail.com}
\author{Pawan Goyal}
\affiliation{
  \institution{Indian Institute of Technology Kharagpur}
  \department{Department of Computer Science and Engineering}
  \city{Kharagpur}
  \postcode{721302}
  \country{India}
}
\email{pawang@cse.iitkgp.ernet.in}

\author{Animesh Mukherjee}
\affiliation{
  \institution{Indian Institute of Technology Kharagpur}
  \department{Department of Computer Science and Engineering}
  \city{Kharagpur}
  \postcode{721302}
  \country{India}
}
\email{animeshm@cse.iitkgp.ernet.in}

\thanks{This research has been performed when SKM and AC were at IIT Kharagpur, India}
\begin{abstract}
In Twitter, there is a rising trend in abusive behavior which often leads to incivility\footnote{\label{ff}https://phys.org/news/2016-11-twitter-tool-curb-online-abuse.html}. This trend is affecting users mentally and as a result they tend to leave Twitter and other such social networking sites thus depleting the active user base. In this paper, we study factors associated with incivility. We observe that the act of incivility is \textit{highly correlated with the opinion differences between the account holder (i.e., the user writing the incivil tweet) and the target (i.e., the user for whom the incivil tweet is meant for or targeted), toward a named entity}. We introduce a character level CNN model and incorporate the \textit{entity-specific sentiment  information} for efficient incivility detection which significantly outperforms multiple baseline methods achieving an impressive  accuracy of \textcolor{black}{\textbf{93.3\%}} (\textcolor{black}{\textbf{4.9\%}} improvement over the best baseline). In a post-hoc analysis, we also study the behavioral aspects of the targets and account holders and try to understand the reasons behind the incivility incidents. Interestingly, we observe that there are strong signals of repetitions in incivil behavior. In particular, we find that there are a significant fraction of account holders who act as repeat offenders - attacking the targets even more than 10 times. Similarly, there are also targets who get targeted multiple times. In general, the targets are found to have higher \textit{reputation scores} than the account holders.
\end{abstract}

%
%


%
%


\maketitle

\section{Introduction}
Twitter is one of the most popular online micro-blogging and social networking platforms. However, of late, this social networking site has turned into a destination where people massively abuse and act in an incivil manner. This trend is affecting the users
mentally and as a result they tend to leave Twitter and other such social networking sites. In an article in Harvard Business Review 2016\cite{hbr}, the author claims that abuse and bullying are the primary reasons why this micro-blogging site is losing its active user base. For instance, in August 2014, the Gamergate controversy\footnote{https://en.wikipedia.org/wiki/Gamergate\_controversy}~\cite{massanari2017gamergate,chatzakou2017measuring,chatzakou2017mean,guberman2016quantifying} broke out in Twitter and other social media platforms. The Gamergate controversy originated from alleged improprieties in video game journalism and quickly grew into a larger campaign (conducted primarily through the use of the hashtag \#GamerGate) centered around
sexism and social justice. Gamergate supporters took to massive incivil behavior e.g., sexual harassment, doxing, threats of rape and murder.

\subsection{What is incivility?}
In general, incivility involves sending harassing or threatening messages (via text message or e-mail), posting derogatory comments about someone on a website or a social networking site (such as Facebook, Twitter etc.), or physically threatening or intimidating someone in a variety of online settings~\cite{grigg2010cyber,corcoran,burgess2009cyberbullying,lenhart2007cyberbullying,li2007new,patchin2006bullies}. In contrast, cyberbullying is defined in the literature as intentional incivil behavior that is repeatedly carried out in an online context against a person who cannot easily defend himself or herself~\cite{kowalski2012cyberbullying,patchin2012update,ybarra2012defining}. The important differences between incivility and cyberbullying are that for the latter there is i) a continuous repetition of the act and ii) the existence of imbalance of power between the target and the perpetrator.

\subsection{Impact of incivility/bullying}
Since online content spread fast and have a wider audience, the persistence and durability can make the target as well as bystanders read the account holder's words over and over again resulting in strongly adverse effects~\cite{campbell2005cyber}. It can, thereby, potentially cause devastating psychological setbacks like depression, low self-esteem, suicide ideation, and may even ultimately lead to actual suicides among the targets~\cite{mcnameebully} ~\cite{menesini2009cyberbullying,hinduja2010bullying}. Teenagers are mostly affected by this; in fact, more than half of the American teens have been the targets of incivility/bullying\footnote{https://cyberbullying.org/}. Not only kids, adults too are subjected to incivility/bullying\footnote{https://cyberbullying.org/bullying-is-not-just-a-kid-problem}, \footnote{https://cyberbullying.org/Preventing-cyberbullying-top-ten-tips-for-adults.pdf}. Owing to the increasing prevalence of social media, celebrity bullying also takes place frequently which is supported by various articles ~\cite{blogxilla, danbully,noellebully}. Facebook, Twitter, YouTube, Ask.fm, and Instagram have been listed as the top five networks having the highest percentage of users who report incidents of incivility\footnote{https://www.ditchthelabel.org/research-papers/}. 

The rapid spread of this activity over the social media calls for immediate active research to understand how incivility occurs on OSNs today ~\cite{sharonharass}. This investigation can help in developing effective techniques\footref{ff} to accurately detect and contain cases of incivility.

\subsection{Working definition of incivility used in this paper}
In this paper, we are specifically interested to study in depth the incivil behavior of Twitter users. Note that we do not consider cyberbullying in this study since it is very different in characteristics from incivility and calls for a completely separate line of investigation.

In the following we present a working definition of incivility that is largely accepted in the community and shall be used all through in the rest of the paper. 

\noindent\textbf{Working definition}: We adopt the definition of incivility as an \emph{act of sending or posting mean text messages} intended to mentally hurt, embarrass or humiliate another person using computers, cell phones, and other electronic devices~\cite{grigg2010cyber,corcoran,dinakar2011modeling,dinakar2012common,singh2017they}. In the dataset section we precisely operationalize this definition for the experiments that follow.
\subsection{Research objectives and contributions}
In this paper, we analyze a large Twitter dataset for incivility detection followed by a detailed post-hoc analysis of the detected tweets. Toward this objective, we make the following contributions.
\begin{compactitem}
 \item We study the behavioral aspects of the targets and account holders and try to understand the reason for the incivility incidents. Our central observation is that \textit{incivility in Twitter is strongly correlated to opinion conflicts between the account holder and the target}.  Further analysis of target and account holder profiles across the linguistic and the cognitive dimensions reveals that account holders generally tend to use \textit{more swear words},\textit{ negations}; express \textit{more emotions} and tweet more related to \textit{body} and \textit{sexual} categories. 
\item Once we have established the association of opinion differences between account holder and target toward named entities in incivility incidents, we propose a deep learning framework based on bi-directional LSTMs and character-CNNs and incorporate the entity-specific sentiment and followership information. Our model achieves an accuracy of \textcolor{black}{\textbf{93.3\%}} with an F1-score of \textcolor{black}{\textbf{0.82}} which significantly outperforms (\textcolor{black}{\textbf{4.9\%}}, \textcolor{black}{\textbf{6.5\%}} improvements w.r.t accuracy, F1-score respectively) the best performing baseline. 
\item We then conduct a post-hoc analysis of the incivility tweets on the entire dataset and study \textit{repetitions in incivility}. We find that there exists \textit{a significant fraction of account holders who act as repeat offenders}. In line of previous works~\cite{pavlopoulos2017deeper,elsherief2018peer,elsherief2018hate}, our results also confirm prior findings of targets being attacked multiple times and targets having usually more followers. The existence of this \textit{imbalance of power} in terms of the social prestige and reputation between the target and the account holder is quite interesting because of its role reversal unlike cyberbullying incidents where the power lies mostly with the account holder.
\end{compactitem}
\vspace{-2mm}

\subsection{Outline of the paper}

\begin{figure}[h]
\vspace{-4mm}
\centering
\includegraphics[scale=0.4]{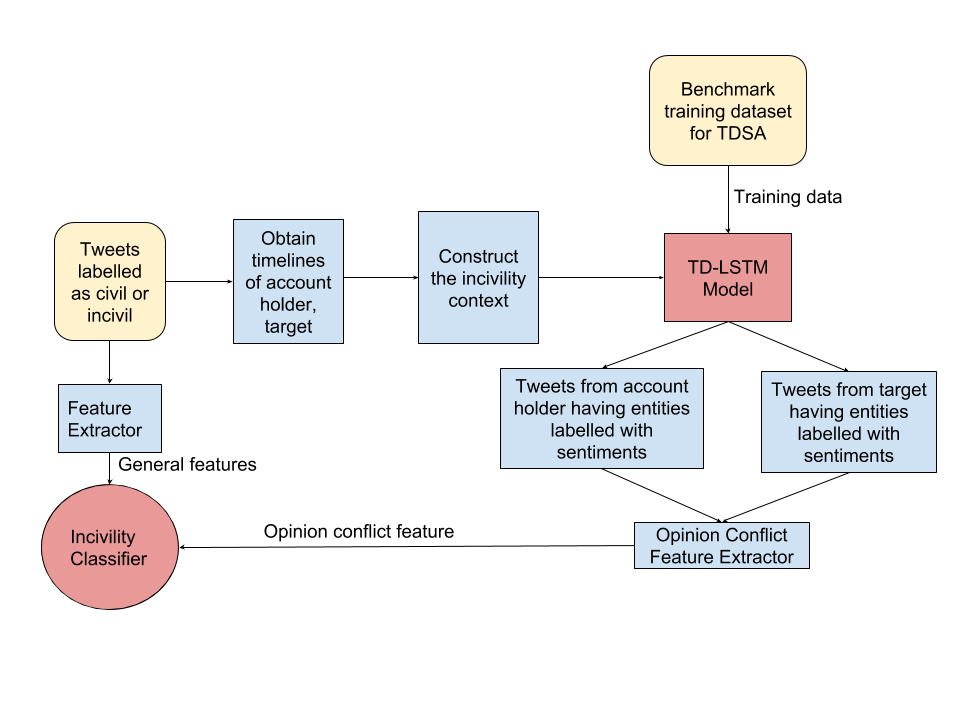}
\caption{Schematic of the steps for incivility detection. The yellow colored blocks represent inputs, the red colored blocks represent the classifiers and the blue colored blocks represent the intermediate steps.}
\label{fig:flowchart}
\vspace{-4mm}
 \end{figure}

We have various subsegments in this paper that we build up layer by layer to finally obtain the detection model. The entire pipeline is illustrated in Figure~\ref{fig:flowchart}.  As a first step tweets are labeled as incivil and then the timelines of the account holder and the target are crawled. This is followed by the construction of the \textit{incivility context} (to be defined later). From the context, target sentiments are extracted using standard Target Dependent Sentiment Analysis (TDSA) technique. Opinions of the account holders and the targets are compared to finally ascertain opinion conflicts. This feature is then pipelined into to a deep neural framework to automatically classify a tweet as incivil.

\section{Related Work}
There have been several works in the area of incivility - online harassment, cyberbullying, trolling etc., by researchers from various communities encompassing sociologists, psychologists and computer scientists. The overall literature in this area can be classified into two major categories based on the approaches taken to address the issues in incivility -- i) survey/interview based approaches (non-computational) ii) computational approaches.

\subsection{Survey/interview based approaches}
Some of the very early works in the area of cyberbullying came from developmental psychology and sociology using survey/interview based approaches. Most of these studies deal with cyberbullying among school students and youth. \\
\noindent\textit{\textbf{Prevalence of bullying}}: The prevalence of cyberbullying incidents in these studies varies from $\sim6\%$ to 72\%. Finkelhor et al.~\cite{finkelhor2000online} report that $\sim6\%$ of youth have been harassed online whereas Juvonen and Gross~\cite{juvonen2008extending} report this number to be 72\%. Most of the other studies estimate that $6-30\%$ of teens have experienced cyberbullying~\cite{hinduja2012}. Similarly, the number of youth who admit to cyberbullying is relatively lesser (3-44\%)~\cite{hinduja2012}. These variability in the numbers is largely due to the following facts -- cyberbullying has been defined in various different ways~\cite{vandebosch2008defining,menesini2009cyberbullying,ybarra2012defining} and also sometimes the perception of bullying differs across age (youth understand it differently compared to adults~\cite{boyd2014s,marwick2011drama}); the authors in these studies adopted various different sampling and methodological strategies~\cite{tokunaga2010following}. Despite differences in rate of cyberbully incidents, the prominence of the factor among adolescents is well-grounded. 

\noindent\textit{\textbf{Sociodemographics}}: Cyberbullying though similar to traditional (offline) bullying embarks characteristic differences in sociodemographics~\cite{schneider2012cyberbullying}. In contrast to school bullying (offline) where boys are found to be more likely targets~\cite{carlyle2007demographic,nansel2001bullying}, in cyberbullying there is no consensus on the role of gender in bullying~\cite{tokunaga2010following}. Some studies have found that girls are more likely to be targets of cyberbullying~\cite{kowalski2007electronic,wang2009school} yet other studies have found no gender differences~\cite{williams2007prevalence,hinduja2008,ybarra2004youth}. Age is another characteristic in which
cyberbullying patterns differ from traditional bullying. Although there is a decreasing trend of traditional bullying from middle to high school~\cite{nansel2001bullying}, some studies suggest that cyberbullying incidents increase during the middle school years~\cite{williams2007prevalence,kowalski2007electronic} and others have found no
consistent relationship between cyberbullying and age~\cite{juvonen2008extending,smith2008cyberbullying}. 

\noindent\textit{\textbf{Effect of cyberbullying}}: There have been studies on the effect of cyberbullying incidents on the adolescents. Researchers have found that cyberbullying incidents are associated with experience of sadness, anger, frustration, depression, embarrassment or fear~\cite{hinduja2007offline,ybarra2007prevalence,patchin2011traditional,mishna2010cyber} and these
emotions have been correlated with delinquency and interpersonal
violence among youth and young adults~\cite{aseltine2000life,broidy1997gender,mazerolle2000strain,mazerolle1998linking}. Apart from these, cyberbullying has been associated with various other behavioral and psychological outcomes like suicide, school drop-out, aggression and fighting, drug usage and carrying weapons to schools~\cite{ericson2001addressing,hinduja2007offline,hinduja2008,hinduja2010bullying,hinduja2015,ybarra2004youth,ybarra2007prevalence,seals2003bullying}. In a recent study by Singh et al.~\cite{singh2017they}, the authors have observed the influence of newer mobile app features like perceived ephemerality, location-based communication and image-based messaging etc. on cyberbullying in high school and its effect on school students.

\noindent\textbf{Critical observations from the above}: Interpretation of cyberbullying seems to be conditioned on age (youths and adults interpret them differently). The role of gender in cyberbullying is unclear. There could be a multitude of effects of cyberbullying ranging from depression, sadness, embarrassment to more severe incidents like fighting, drug use and carrying weapons. 
\subsection{Computational approaches} 
There have been several works in computer science domain mainly focusing on automatic detection of cyberbullying in social media largely based on text analytic approaches applied to online comments~\cite{yin2009detection,sanchez2011twitter,reynolds2011using,dinakar2012common,xu2012learning,chen2012detecting,kontostathis2013detecting}. 
\\
\noindent\textit{\textbf{Content based approaches}}: \\
\textbf{Language usage}: There are several works that focus on usage of language - specific high frequent terms associated with bullying incidents. Reynolds et al.~\cite{reynolds2011using} use curse and insult words with their level of offensive intensity as primary indicators for detection of cyberbully incidents. Chen et al.~\cite{chen2012detecting} propose a user-level offensiveness detection method by introducing Lexical Syntactic Feature (LSF) architecture to detect offensive content and potential offensive users. They observe the contribution of pejoratives/profanities and obscenities in determining offensive content, and propose hand-crafted syntactic rules to detect name-calling harassments. They incorporate user's writing style, structure and specific cyberbullying content as features to predict the user's potential to post offensive content. Kontostathis et al.~\cite{kontostathis2013detecting} also focus on language used in cyberbullying incidents. They perform analysis of the words used in connection to cyberbullying incidents on Formspring.me and further use those words and their context words to build a cyberbullying detection framework.

\noindent\textbf{Gender role}: Chisholm in a social study~\cite{chisholm2006cyberspace} show that there exist differences between males and females in the way they bully each other. Females tend to use relational styles of aggression, such as excluding someone from a group and ganging up against them, whereas males use more threatening expressions and profane words. Following this study, Dadvar et al.~\shortcite{dadvar2012improved} leverage gender-specific language usage for cyberbully detection on MySpace profiles and show that such gender information improves the accuracy of the classifier.  

\noindent \textbf{Sentiment usage}: Hee et al.~\cite{van2015automatic} use sentiment lexicon and content of the text as bag-of-words (unigram/bigrams of words and character trigrams) for detection of cyberbully posts on a corpora of $\sim91,000$ Dutch posts from Ask.fm. Nahar et al.~\cite{nahar2012sentiment} use sentiment features generated from Probabilistic Latent Semantic Analysis (PLSA) on cyberbully texts and bag-of-word features to detect cyberbullying incidents. Further, they detect and rank the most influential persons (bullies and targets).

\noindent \textbf{Mitigation of cyberbullying incidents}: Dinakar et al.~\shortcite{dinakar2012common} address the problem of detection of cyberbullying incidents and propose an intervention technique by notifying participants and network moderators and offering targeted educational material. In this work, they present an approach for bullying detection based on natural language processing and a common sense knowledge base that allows recognition over a broad spectrum of topics in everyday life. They construct BullySpace, a common
sense knowledge base that encodes particular knowledge about bullying from various associated subject matters (e.g., appearance,
intelligence, racial and ethnic slurs, social acceptance, and rejection) and then perform a joint
reasoning with common sense knowledge. To mitigate the problem of cyberbullying, they propose a set of intervention techniques. They propose an ``air traffic control''-like dashboard, that alerts moderators to large-scale outbreaks of bullying incidents that appear to be escalating or spreading and help them prioritize the current deluge of user complaints. For potential victims, they provide educational material that informs them about coping with the situation, and connects them with emotional support from others. 

\noindent\textit{\textbf{Leveraging contextual information}}: \\
Apart from the content of the text, the contextual information is also important and relevant for cyberbully detection since the lexicon based filtering approach is prone to problems around
word variations and lack of context. Yin et al.~\shortcite{yin2009detection} use a supervised learning methodology for cyberbully detection using content and sentiment features, as well as contextual (documents in the vicinity) features of the considered documents on Slashdot and MySpace dataset. Similarly, Zhong et al.~\shortcite{zhongcontent} study cyberbullying of images on Instagram using text as well as image contextual features. Hosseinmardi et al.~\shortcite{li2014comparison} examine the users who are common to both Instagram and Ask.fm and analyze the negativity and positivity of word usage in posts by common users of these two social networks. Hosseinmardi et al. in two related works~\cite{hosseinmardi2015detection,hosseinmardi2015prediction} study detection and prediction of cyberbullying. In~\cite{hosseinmardi2015detection}, the authors detect cyberbullying in Instagram by classifying images to different categories and further including text features from comments. In~\cite{hosseinmardi2015prediction}, the authors try to predict cyberbullying instances using the initial posting of the media object, any image features derived from the object, and any properties extant at the time of posting, such as graph features of the profile owner. In a recent paper~\cite{chatzakou2017mean}, Chatzakou et al. employ a more robust approach considering text, user and users' follower-followee network-based attributes for detecting aggression and bullying behavior on Twitter. In this paper, the prime objective of the authors is to distinguish cases of cyberbullying from aggression and spamming. In a related vein, Chatzakou et al.~\cite{chatzakou2017measuring,chatzakou2017hate} study cyberbullying and aggression behavior in GamerGate controversy (a coordinated campaign of harassment in the online world) on Twitter. 

Kwak et al.~\cite{Kwak:2015} have analyzed a large-scale dataset of over 10 million player reports on 1.46 million toxic players from one of the most popular online game in the world, the League of Legends. They observe reporting behavior and find that players are not engaged in actively reporting
toxic behavior and this engagement can be significantly improved via explicit pleas from other players to report. There are significantly varying perceptions of what constitutes toxic behavior between those that experienced it and neutral third parties. There are biases with respect to reporting allies vs. enemies. There are also significant cultural differences in perceptions
concerning toxic behavior.

Chen et al.~\cite{chen2017presenting} in a recent study, present a dataset of user comments,
using crowdsourcing for labeling. Due to ambiguity and subjectivity in abusive content from the perspective of individual reader, they propose an aggregated mechanism for assessing different opinions from different labelers. In addition, instead of the typical binary categories of abusive or not, they introduce an additional third category of `undecidedness' to capture
the instances that are neither blatantly abusive nor clearly harmless. They evaluate against the performance of various feature groups, e.g., syntactic, semantic and context-based features which yield better classification accuracy. Samghabadi et al.~\cite{samghabadi2017detecting} perform similar study of nastiness (invective in online posts) detection in social media. They present evolving approaches for creating a linguistic resource to investigate nastiness in social media. The starting point is selecting profanity-laden posts as a likely hostile source for invective potentially leading to cyberbullying events. They use various types of classic and new features, and try to combine them for distinguishing extremely negative/nasty text from the rest of them.
In a recent study, Hosseini et al.~\cite{hosseini2017deceiving} analyze the recent advancement of Google's Perspective API for detecting toxic comments and show that the system can be fooled by slight perturbation of abusive phrases to receive very low toxicity scores, while preserving the intended meaning. 

Pavlopoulos et al.~\cite{pavlopoulos2017deeper} have recently introduced deep neural models to moderate abusive (hate speech, cyberbullying etc.) user content.

\noindent\textbf{Critical observations from the above}: Both content and context have been extensively used to design hand-crafted features for detection of cyberbullying. There have also been one or two attempts to use deep learning techniques to abusive content moderation in general. However, there are hardly any approach that marry deep neural models with features that could be critically responsible for the invocation of incivil posts.

\noindent \textbf{Hate Speech}: Hate speech detection has been studied by various researchers. These works use several lexical properties such as n-gram features~\cite{nobata2016abusive}, character n-gram features~\cite{mehdad2016characters}, word and paragraph embeddings~\cite{nobata2016abusive,djuric2015hate} to detect hate speech. Apart from detection, there exist research works that look into various aspects of hate targets and the instigators.  Silva et al.~\cite{silva2016analyzing} study the targets of online hate speech by searching for sentence structures similar to ``I <intensity> hate <targeted group>''. They find that the top targeted groups are primarily bullied for their ethnicity, behavior, physical characteristics, sexual orientation, class, or gender. ElSherief et al.~\cite{elsherief2018peer} present the comparative study of hate speech instigators and target users on Twitter. They study the characteristics of hate instigators and targets in terms of their profile self-presentation, activities, and online visibility and observe that hate instigators target more popular (celebrities with a higher number of followers) Twitter users. Their personality analysis of hate instigators and targets show that both groups have eccentric personality facets that differ from the general Twitter population. In another concurrent paper, ElSherief et al.~\cite{elsherief2018hate} focus on the hate targets - either directed toward an individual or toward a group of people. They perform the linguistic and psycholinguistic analysis of these two forms of hate speech and show that directed hate speech, being more personal and directed, is more informal, angrier, and often explicitly attacks the target (name calling) with fewer analytic words and more words suggesting authority and influence. Generalized hate speech, on the other hand, is dominated by religious hate, is characterized by the mentions of lethal words such as murder, exterminate, and kill; and quantity words such as million and many. In our study, as well, we observe similar findings regarding the accounts and targets.

\noindent \textbf{Trolling:} Trolling has been defined in the literature as behavior that falls outside acceptable bounds defined by those communities~\cite{binns2012don,hardaker2010trolling}. There have been divided opinions on trolling behavior. Prior research works suggest that trolls are born and not made: those engaging in trolling behavior have unique personality traits~\cite{buckels2014trolls} and motivations~\cite{baker2001moral,herring2002searching,shachaf2010beyond}. However, there is other school of thought suggesting that people can be influenced by their environment to act aggressively~\cite{cialdini2004social,jones1978air}. Cheng et al.~\cite{cheng2017anyone} in a recent study, focus on the causes of trolling behavior in discussion communities. By understanding the reason behind trolling and its spreads in communities, one can design more robust social systems that can guard against such undesirable behavior.

\subsection{The present work}
Our work is different from the previous works in several ways. We focus on understanding the incivility incidents on general population in Twitter (unlike most of the previous studies which are based on children or teens and dated online platforms like FormSpring, MySpace etc.). While most of the survey/interview based techniques concentrate on analyzing the effects and consequences of these incidents after the incident has happened, we attempt to (i) early detect the incidents of incivility in the first place and then (ii) analyze the behavioral aspects of the account holder and the targets so detected. Though there have been several lexicon-based studies, they have rarely reused any labeled data from previous researchers due to incompatibility issues of applying on different platforms. Our study, though on Twitter, can be used in other social media platforms because of presence of a single user in multiple platforms and essentially similar language usage trend.

In particular, we observe that opinion conflict (target and account holder showing opposite sentiments towards same named entities) is associated with incivility and we believe that this aspect has not been earlier reported in the literature, not even in any of the earlier computational studies. We then propose a deep learning based model that can automatically detect incivility text more efficiently by correctly exploiting the aforementioned connection between incivility incidents and target dependent expression of sentiments. This fusion of a deep neural model with a critical feature that stems from opinion conflict is a prime novelty of our work.

\section{Dataset preparation}\label{dataset}
Recall that our working definition of incivility refers to the act of sending or posting mean text messages intended to mentally hurt or embarrass a target. In this section we describe how we operationalize this definition through the assemblage of appropriate data.

\subsection{Operationalizing the definition of incivility} Although incivility is rising in Twitter\footref{ff}, from automated crawls, it is difficult to obtain data containing direct incivility instances. However, our working definition presents us with one important clue that incivility tweets should generally contain \emph{offensive/mean} words. We use a list of such offensive words compiled by Luis von Ahn of Carnegie Mellon University\footnote{https://www.cs.cmu.edu/~biglou/resources/bad-words.txt}. This list is very comprehensive and larger than any list containing offensive words known to us or used previously by other authors. This list has been used by various authors~\cite{Fredheim:2015,Beelen,chen2017presenting,addawood2017telling,bruni2014multimodal,phan2017play} in the past and this rich literature justifies our choice. The list contains 1374 offensive words. The offensive words included in the list can be categorized as: (i) Swear/profane words: \emph{f**k}, \emph{a\$\$hole}, \emph{b*tch} etc.,
(ii) Negative words: \emph{die}, \emph{hell}, \emph{death} etc. and (iii) Others: \emph{enemy}, \emph{drug} etc.

Note that the list contains certain words that some of the people would not find offensive. However, incivil tweets have a very high chance of having these offensive words\footnote{Note that the number of incivil tweets in a random sample of tweets is very low. In fact, we observe that in a random sample of 100 tweets, there are only 2 tweets that fall into the category of incivility. Also the tweets that were incivil from the random sample contained words from the ``mean'' words list.  After obtaining  manually annotating the tweets as incivil or civil , we found that the fraction of mean words were higher in the incivil tweets.}. We do understand that these may not cover all instances of incivility and might affect the recall of our models which is a limitation of our treatment of the problem. However, the precision of the system should not be affected by this limitation and the current target of the paper is to build a highly precise system. 

We then use these words to filter the Twitter data stream. The crawling has been done from August to December 2017.
We filter the tweets in which any of those offensive words are present. 
We obtain a total of $\sim$2,000,000 tweets containing one or more such offensive words. Out of the 1374 offensive words, 59.4\% of words have been used at least once, 54.9\% have been used at least twice and 47.8\% have been used at least thrice. Some offensive words have also been used in as high as 1.3\% of the tweets.

%

\subsection{Mention based filtering of the dataset}
We further filter the dataset based on the presence of mentions in the tweets. This is because in general, any conversation in Twitter should contain mention(s) and an account holder would generally mention the target in their (incivil) tweets. We consider only those tweets in which one or more mentions appear. This reduces the tweet dataset to $\sim$300,000 tweets from the earlier $\sim$2,000,000 tweets. Note that this step operationalizes the second part of the working definition of incivility, i.e., the offensive text is usually targeted to an individual, which in this case is the mentioned target.

\subsection{Manual labeling of the tweets}
While the previous step reduces some noise in the data, not all the tweets obtained are related to incivility. We, therefore, randomly sample 25,000 tweets out of these $\sim$300,000 tweets and manually label them as incivil/civil instances. We consider those cases where we have full agreement between the two authors who did the labeling. Out of the 25,000 tweets, we discard 729 tweets where there was a disagreement between the authors. In the 24,271 selected tweets where there was full agreement, we find 8,800 tweets as instances of incivility and the remaining 15,441 tweets as instances of civility.

\subsection{Criteria for labeling the tweets}	
Following are some of the broad cases where we have labelled the tweets as being incivil:\\ \textbf{(a) \textit{Blackmails or threats}:} These tweets have expressions of physical or psychological threats to the targets. Example: `I'll smash you in the face when I see you.' \\ \textbf{(b) \textit{Insult}:} These tweets have insults that are abusive for the target. Example: `You are such an a\$\$hole.'\\
\textbf{(c) \textit{Cursing}:} These tweets have expressions wishing that some grave misfortune befalls the target (like their or their loved one's death). Example: `You'll die and burn in hell.' \\
\textbf{(d) \textit{Sexual harassment}:} These tweets contain unwanted sexual talk which might be derogatory. Example: `Post a naked pic u sl*t!'

While labelling the tweets, we took into account that some of these tweets although might contain offensive words are not cases of incivility, eg., `Hey bitches, feel like seeing a movie tonight?'

\section{Target sentiments \& incivility}
In this section, we construct an \textit{incivility context} to be able to perform an in-depth analysis of the reason for incivility. We find that contradicting sentiments (opinion conflicts) by the account holder and target toward a named entity, are highly correlated to the act of incivility.
\subsection{The incivility context}
As described in the previous section, we label the incivil tweets and from the mention relationship, identify the target and the account holder. A typical example is as follows: 
\begin{tcolorbox}[boxsep=0pt, top=1pt,left=1pt,right=1pt,bottom=1pt,colframe=gray!50]{\footnotesize
\noindent @user1 what a dumb ass ** person you are, user2

\noindent Here \emph{user2} is the account holder while \emph{user1} is a target.}
\end{tcolorbox}

In our work, we consider the mentions of an incivil tweet as the target. In case of multiple mentions, we consider all of the mentions as targets. However, it might be possible that some of the mentions are not actually targets (usually rare as observed through manual inspection) and is currently a limitation of our work which we wish to address in the future.
Once we have the account holder-target pairs, we consider the targets' and the account holders' profiles and crawl their timeline (we get a maximum of 3,200 tweets per user). Also note that the timeline is crawled at around the same time we collected our dataset. We take the first 100 tweets from their timeline. We find that taking the first 100 tweets itself contains ample clues for us to investigate.
The intuition behind constructing this context is that these tweets might contain the context that triggered the incivil tweet and thus these tweets should provide the best clues as to why the account holder targeted the target. In fact, for annotating the tweets in the previous section, the annotators in many instances had to visit the target and the account holder profiles to ensure if the instance was indeed a case of incivility. In almost all the cases where the annotators attested instances of incivility they observed opinion conflicts between the target and the account holder. We therefore attempt to automate this natural ``back-and-forth'' process adopted by the annotators for identifying incivility instances and build the incivility context to automatically analyze such opinion conflicts by making comparisons between the target's and the account holder's timelines. 

From the example of incivility context cited below, we observe that the target tweets positively about \emph{Donald Trump} and \emph{US Economy}. However, the account holder tweets negatively about \emph{Trump} and positively about \emph{President Obama}. We can observe that there is a conflict of opinion between the target and the account holder as the sentiments expressed toward the common named entity Donald Trump is opposite. Going through the entire context, we find that this opinion conflict leads to an incident of incivil post.
\begin{tcolorbox}[boxsep=0pt, top=1pt,left=1pt,right=1pt,bottom=1pt,colframe=gray!50]{\footnotesize{
\noindent \textbf{account holder's tweet}: \\
\noindent @user1 Enjoy prison a\$\$hole!  \\
\noindent \textbf{account holder's context tweets}:\\
\noindent @user5 @user6 You sir, are just another clueless Trump lemming.\\
\noindent @user7 @user8 Seriously, get your head out of Trump's ass already. Go watch your Fox News \& Friends and eat your jello.\\
\noindent @user8 The video of what your boyfriend said: Trump labels US justice system 'laughingstock' @CNNPolitics https://t.co/QNa2jqAYsE\\
\noindent @user9 if the Devil was running as a Republican, would you still vote for him? Your morals and priorities are so screwed up.\\
\noindent @user5 Seriously, let it f**king go. You are worse that a scorned girlfriend bringing up decades of shit that does not matter. You are the BIGGEST LOSER of all time.\\
\noindent @user11 Trump idiot lemmings are condemning the outrage over slavery and agreeing w/the idiot Kelly about praising Lee? Clueless losers\\
\noindent \textbf{target's context tweets}:\\
\noindent @user10 You are truly stupid. Trump is the first President to come into Office supporting marriage equality\\
\noindent Strange that the \#fakenews media never gets stories wrong in favor of Trump. It's almost like they do it on purpose\\
\noindent According to HuffPo, President Trump is effective, but they don't like it.
\noindent Donald Trump's relentless focus on tax cuts, deregulation and draining the swamp is great for job growth... with minorities\\

\noindent and so on ...}
}
\end{tcolorbox}
Based on the 8800 incivility incidents in our manually labeled dataset, we obtain the incivility contexts. Since our hypothesis is that opinion conflicts between the account holders and the targets create the rift between them, we plan to use sentiment analysis as a tool to automatically identify the conflicts. In particular, we try to find out the sentiments expressed by the targets toward different named entities in their tweets. We also find out the sentiments expressed by the account holders toward the different named entities. Next we find the common named entities toward which the target and the account holder expresses positive or negative sentiments. Finally, we identify if the targets and the account holder have expressed opposing sentiments toward the same entity. 
We propose an algorithm to detect such opinion conflicts easily in the following subsections. 

\subsection{Step 1: Named entity recognition}
We use a named entity recognition (NER) system to identify named entities that will enable us to proceed further with the sentiment analysis. Recall that, according to our hypothesis, we first need to identify the named entities and then find the sentiments expressed by the users toward these named entities. For this purpose, we use the tool proposed in~\cite{Ritter11,Ritter12}\footnote{https://github.com/aritter/twitter\_nlp} that is trained on Twitter data and performs much better than other traditional NER tools like the ones proposed in ~\cite{Finkel}\footnote{http://nlp.stanford.edu/software/CRF-NER.shtml}.

\subsection{Step 2: Target dependent sentiment analysis}
After named entity recognition, we set out to obtain the sentiment polarity associated with each named entity in a given text. Our sentiment analysis model is based on an LSTM framework. A brief description of LSTMs and how we use it for target sentiment detection follows below. 

\noindent \textit{Long Short Term Memory (LSTM)}
Recurrent Neural Networks (RNNs) are used for handling tasks that use sequences. A recurrent network takes an input vector $x$ and outputs a vector $y$. However, $y$ depends on not only the current input, but on all the inputs fed into it in the past. RNNs are however not capable of handling long term dependences in practice \cite{bengio1994learning}. 
It was observed that the backpropagation dynamics caused gradients in an RNN to either vanish or explode. The exploding gradient problem can be solved by the use of gradient clipping. \cite{hochreiter1997long} 
introduced LSTM to mitigate the problem of vanishing gradient. The LSTMs by design have a hidden state vector ($h_t$) and also a memory vector $c_t$ at each timestep $t$. The gate equations at timestep $t$ with input as $x_t$ and output as $o_t$ are:
\[ f_t = \sigma({W_f}.[h_{t-1}, x_t] + b_f) \]
\[ i_t = \sigma({W_i}.[h_{t-1}, x_t] + b_i) \]
\[ C_t = \tanh({W_C}.[h_{t-1}, x_t] + b_C) \]
\[ c_t = f_t \circ c_{t-1} + i_t \circ C_t\]
\[ h_t = o_t \circ \tanh (c_t)\]
\[ o_t = \sigma({W_o}.[h_{t-1}, x_t] + b_o)\] 
Here, the three vectors $f, i, o$ are thought of as binary gates to control whether each memory cell is updated.
The $\circ$ operation here denotes the element-wise matrix multiplication.
\

\noindent\textit{Target dependent sentiment classification using LSTM}: Here we describe a LSTM based model for target dependent (TD) sentiment classification. The model we use is adapted from~\cite{tang2015effective}. The sentence representation in this model can be naturally considered as the feature to predict the sentiment polarity of the sentence. We use the glove vector word embeddings which are trained on $27$ billion tokens and $2$ billion tweets as the vector representation of the words. We use the target dependent LSTM approach (TD-LSTM) to find the sentiment polarity of the tweets toward a target. 

The training of the TD-LSTM is done in a supervised learning framework using cross entropy loss.
%
%
%
We perform all parameter updates through backpropagation and \emph{stochastic gradient descent}.


\noindent\textit{Evaluation results:}
We use the benchmark dataset from~\cite{dong2014adaptive}. We train it on their Twitter dataset having 6248 sentences and test it on 692 sentences. The percentages of positive, negative and neutral instances in both the training and test sets are 25\%, 25\% and 50\% respectively. The accuracies reported in table~\ref{TD-LSTM} are obtained using 200 dimensional glove word embeddings. We manually label the sentiments of 100 random entities from our dataset. We achieve an accuracy of 73.4\%  on our dataset.


\begin{table}
\begin{center}
\small
 \begin{tabular}{|c |c|c|} 
 \hline
 Classifier & Accuracy & Macro F1\\ [0.5ex] 
 \hline
 TD-LSTM on benchmark dataset & 69.2\% & 0.68 \\
 TD-LSTM on our dataset & 73.4\% & 0.71 \\
 \hline
\end{tabular}
\end{center}
\caption{\label{TD-LSTM} Accuracy of the TD-LSTM.}
\vspace{-5mm}
\end{table}

The above accuracy has been obtained using the early stopping criteria and the same weight initializations as mentioned in~\cite{tang2015effective}. 

\subsection{Observations from TD sentiment analysis}
In this section, we analyze the outcome of the TD sentiment analysis on the named entities. For each incivility context, we perform the NER and target dependent sentiment analysis. Specifically, we take all tweets in the incivility context of the account holders and targets and then run the NER on the tweets to extract the named entities. Then we run the TD-LSTM with targets as these named entities to get the target's polarities toward the named entities. We can conclude that the cases of incivility where the algorithm returns empty sets for both the positive and the negative sentiments for all the named entities are not related to contradicting sentiments between the account holder and the target. 

\noindent\textit{Sentiment expression toward different named entities}: 
\begin{figure*}[ht]
\begin{minipage}{0.3\textwidth}
 \centering
\pgfplotstableread[row sep=\\,col sep=&]{
    NE & neu & pos & neg\\
    company     & 0.012 & 0.018 & 0.03\\
    facility & 0.05 & 0.012 & 0.008\\
    other     & 0.019 & 0.048 & 0.039\\
    movie & 0.007 & 0.022 &0.020\\
    band & 0.010 & 0.022 & 0.032\\
    person & 0.076 & 0.175 & 0.321\\
    product & 0.005 & 0.014 & 0.011\\
    mentions & 0.864 & 0.686 & 0.536\\
    }\newentdata

\begin{tikzpicture}[scale=0.7]
    \begin{axis}[
            ybar,
            bar width=0.1 cm,
            legend style={at={(0.3,0.99)},
                anchor=north},
            symbolic x coords={company, facility, other, movie, band, person, product, mentions},
            xtick=data,
            xticklabel style={rotate=45, anchor=east},
            ylabel={$\text{Fraction of entities}$ },
        ]
        \addplot table[x=NE,y=neu]{\newentdata};
        \addplot table[x=NE,y=pos]{\newentdata};
        \addplot table[x=NE,y=neg]{\newentdata};
        \legend{Neutral, Positive, Negative}
    \end{axis}
\end{tikzpicture}
\caption{Comparison of sentiment toward different named entity classes.}
\label{fig:NERSentiment}
\end{minipage}
\hfill
\begin{minipage}{0.3\textwidth}
 \centering
\pgfplotstableread[row sep=\\,col sep=&]{
    EN & count \\
    @RealDT     & 160907 \\
    @Youtube & 70145 \\
    Trump     &  67286\\
    @FoxNews & 24143 \\
    America & 21214\\
    Alabama & 21045\\
    Roy Moore & 20865 \\
    @POTUS & 20661\\
    Twitter & 14219\\
    Christmas & 14070\\
    }\newentdata

\begin{tikzpicture}[scale=0.6]
    \begin{axis}[
            ybar,
            bar width=0.3 cm,
            legend style={at={(0.3,0.99)},
                anchor=north},
            symbolic x coords={@RealDT, @Youtube, Trump, @FoxNews, America, Alabama, Roy Moore, @POTUS, Twitter, Christmas},
            xtick=data,
            xticklabel style={rotate=30, anchor=east, align=right},
			ylabel={$\text{No of targets}$ },
        ]
        \addplot table[x=EN,y=count]{\newentdata};
    \end{axis}
\end{tikzpicture}

\caption{Popular entities toward which different targets express sentiments.(@RealDonaldTrump has been shortened as @RealDT.) }
\label{fig:CommonNE}
\end{minipage}
\hfill
\begin{minipage}{0.3\textwidth}
 \centering
\pgfplotstableread[row sep=\\,col sep=&]{
    NE & neu & pos & neg\\
    0-100     & 0.001 & 0. & 0.002\\
    100-1k & 0.008 & 0.005 & 0.007\\
    1k-10k     & 0.076 & 0.072 & 0.101\\
    10k-100k & 0.364 & 0.353 &0.383\\
    100k-1M & 0.316 & 0.334 & 0.318\\
    1M-10M & 0.176 & 0.177 & 0.146\\
    10M-100M & 0.058 & 0.058 & 0.042\\
    }\newentdata

\begin{tikzpicture}[scale=0.7]
    \begin{axis}[
            ybar,
            bar width=0.1 cm,
            legend style={at={(0.2,0.99)},
                anchor=north},
            symbolic x coords={0-100, 100-1k, 1k-10k, 10k-100k, 100k-1M, 1M-10M, 10M-100M},
            xtick=data,
            xticklabel style={rotate=30, anchor=east, align=right},
            ylabel={$\text{Fraction of users}$ },
        ]
        \addplot table[x=NE,y=neu]{\newentdata};
        \addplot table[x=NE,y=pos]{\newentdata};
        \addplot table[x=NE,y=neg]{\newentdata};
        \legend{Neutral, Positive, Negative}
    \end{axis}
\end{tikzpicture}
\caption{Sentiments versus followership counts of targets.}
\label{fig:FollSentcdf}
\end{minipage}
\end{figure*}
In Figure~\ref{fig:NERSentiment}, we show the distribution of sentiments expressed toward different named entity classes by the targets in the incivility context. It is evident from the figure that users express opinions differently about different entities. For example, many users express negative sentiments toward a mention or a person. However, users express no opinion toward the majority of mentions. 
Further, users express negative sentiments more often than positive sentiments. This observation can act as an effective strategy to predict incivility incidents beforehand.

Next we attempt to identify the popular entities toward which targets express sentiments (see Figure~\ref{fig:CommonNE}).
It is evident from the figure that there are more targets expressing opinion about Trump, Youtube and Fox News. These entities might come on top due to the particular dataset used for this work. 

Table~\ref{tab:table2} shows examples where the target is attacked for expressing positive sentiment toward Trump in an example and negative sentiment in another.

\begin{table}[h]
\vspace{-3mm}
 \centering
 \caption{Sentiment toward a common named entity}~\label{tab:table2}
 \resizebox{14cm}{!}{\small{
 \begin{tabular}{|p{7cm}|p{7cm}|}
 \hline
 Positive Sentiment & Negative Sentiment \\ 
 \hline
 \multicolumn{2}{|c|}{Incivil tweets} \\ \hline 
@user1 you really are a stupid dumb **. Do
you know anything. Like trump you are a ** brainless
**. Dumbo & @user2 @user3 @user4 Since when do you have to be a politician to talk to people with respect you stupid moron. You are a fool.  \\ \hline
 \multicolumn{2}{|c|}{Incivility context} \\ \hline
RT @user5: The Washington Post calls out
\#CrookedHillary for what she REALLY is. A PATHOLOGICAL
LIAR! Watch that nose grow! & I don't think he's got a thing to apologize for... He is not a professional politician\\
RT @user6: The Obama/Clinton admin has
failed Americans: ``The Obama economy is trouble for
Hillary Clinton" https://t.co/JBoCmyHj6v 
 &  @user7: Trump is trying to go negative; drive up those numbers so that he \&; Clinton are on the same page https://t.co/RlB00L32v9\\

\hline
\end{tabular}}}
\vspace{-2mm}
\end{table}
\noindent\textit{Sentiments and followership}: We next study whether followership counts and the sentiments expressed by a target have any correlation. Figure~\ref{fig:FollSentcdf} shows the relationship between the sentiment expressed by the users and their number of followers. There is a distinct trend observed here. Users with moderately low followership (100-10K) tend to express more negative sentiments toward the named entities while the users with high followership (100K-10M) show more positive sentiments toward such entities. Usually incivility is associated to negative sentiments. It seems that users with high followership are less likely to be an account holder. In the later sections, we show that this is indeed true.

\noindent \textit{Opinion conflicts.} We have defined opinion conflicts as the case where the target and the account holder express opposing sentiments toward the same entity. For example, suppose the target T expresses 10 sentiments toward entity E, out of which 8 are positive and 2 are negative. Similarly, suppose the account holder expresses 20 sentiments toward entity E, out of which 4 are positive and 16 are negative. Then, the overall sentiment toward the entity E by T is positive, whereas by A is negative. This counts as a single opinion conflict between A and T. We use this mechanism to compute the opinion conflict feature values. 

We find that opinion sentiments are highly correlated to incivility. We observe that 75\% of the incivil tweets in our dataset have at least \textit{one} opinion conflict between the targets and account holders. In contrast, there are only 53\% civil incidents where there is an opinion conflict. Further, for each incivil incident, there are $2.52\pm0.03$ similar sentiments expressed, whereas, for each civil incident, the mean similar sentiments expressed is $4.49\pm0.19$. Thus, it is clear that most civility contexts come with an agreement of opinion (67\% of total sentiments) whereas an incivility context is filled with opinion conflicts (49\% of total sentiments). Therefore, we use this concept as a potential strategy for incivility detection and further use this as a feature in our subsequent models.

\section{Behavior of the account holders and the targets}

%

In this section, we study the socio-linguistic behavior of the targets and the account holders. Specifically, we analyze their user profiles to study followership behavior, tweeting behavior and psycholinguistic dimensions. 

%

\subsection{Followerships}
\begin{figure}[h]
\vspace{-4mm}
\centering
\includegraphics[scale=0.17]{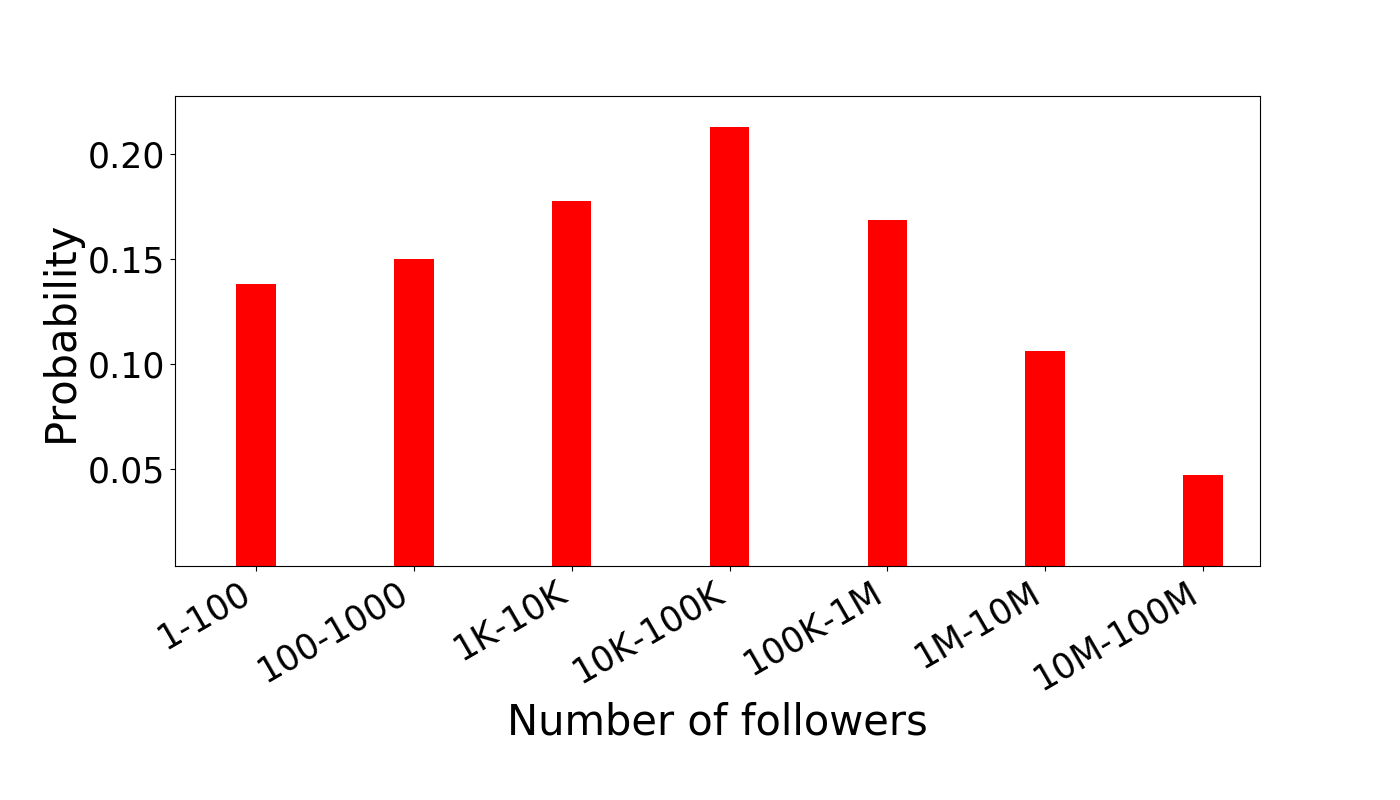}
\caption{Followership distribution of all the users in our dataset.}
\label{fig:followdata}
\vspace{-4mm}
 \end{figure}

In Figure~\ref{fig:followdata}, we show the followership distribution of the user profiles in our dataset. It is evident from the figure that our dataset is a good mix of both normal users as well as celebrity users. Figure~\ref{fig:Followercdf} shows a comparison between the follower counts of the account holder and the target accounts. The figure indicates that the target accounts are more popular than the account holder accounts ($p$-value of significance is  $<10^{-5}$). Apparently, many well known users or celebrities tend to be targets of incivility. A number of news articles also seem to mention this ~\cite{blogxilla,danbully,noellebully}. Moreover, usually celebrities (i.e., those with typically high follower counts) tend to avoid insulting someone on social media sites. 
 
\begin{figure*}[htbp]
    \centering
    \begin{minipage}{.33\textwidth}
        \centering
\pgfplotstableread[row sep=\\,col sep=&]{
    rng & acc & tar \\
    1-100     & 0.05 & 0.013 \\
    100-1k & 0.22 & 0.102 \\
    1k-10k     & 0.49 & 0.34 \\
    10k-100k & 0.215 & 0.297 \\
    100k-1M & 0.022 & 0.155 \\
    1M-10M & 0.002 & 0.067\\
    10M-100M & 0.0001 & 0.019 \\
    }\newentdata

\begin{tikzpicture}[scale=0.7]
    \begin{axis}[
            ybar,
            bar width=0.2 cm,
            legend style={at={(0.7,0.99)},
                anchor=north},
            every tick/.style={
        	black,
        	semithick,
      		},
            symbolic x coords={1-100, 100-1k, 1k-10k, 10k-100k, 100k-1M, 1M-10M, 10M-100M},
            xtick=data,
            xticklabel style={rotate=30, anchor=east, align=right},
            ylabel={$\text{Probability}$ },
            xlabel= No of followers,
        ]
        \addplot table[x=rng,y=acc]{\newentdata};
        \addplot table[x=rng,y=tar]{\newentdata};
        \legend{Account Holder, Target}
    \end{axis}
\end{tikzpicture}
\caption{Comparison of number of followers of the account holders and the target accounts.}
\label{fig:Followercdf}
        \end{minipage}%
        \hfill
    \begin{minipage}{.33\textwidth}
        \centering
\pgfplotstableread[row sep=\\,col sep=&]{
    rng & acc & tar \\
    1-100     & 0.007 & 0.004 \\
    100-1k & 0.041 & 0.024 \\
    1k-10k     & 0.166 & 0.113 \\
    10k-100k & 0.388 & 0.355 \\
    100k-1M & 0.354 & 0.43 \\
    1M-10M & 0.042 & 0.071\\
    }\newentdata

\begin{tikzpicture}[scale=0.7]
    \begin{axis}[
            ybar,
            bar width=0.2 cm,
            legend style={at={(0.18,0.99)},
                anchor=north},
            every tick/.style={
        	black,
        	semithick,
      		},
            symbolic x coords={1-100, 100-1k, 1k-10k, 10k-100k, 100k-1M, 1M-10M},
            xtick=data,
            xticklabel style={rotate=30, anchor=east, align=right},
            ylabel={$\text{Probability}$ },
            xlabel= No of tweets,
        ]
        \addplot table[x=rng,y=acc]{\newentdata};
        \addplot table[x=rng,y=tar]{\newentdata};
        \legend{Account Holder, Target}
    \end{axis}
\end{tikzpicture}
\caption{Comparison of the number of tweets posted by the account holder and the target accounts.}
\label{fig:Statuscdf}
    \end{minipage}%
    \hfill
    \begin{minipage}{.32\textwidth}
        \centering
\pgfplotstableread[row sep=\\,col sep=&]{
    rng & acc & tar \\
    0 & 0.148 & 0.280 \\
    1 & 0.572 & 0.451 \\
    2 & 0.163 & 0.141 \\
    3 & 0.059 & 0.054 \\
    4 & 0.024 & 0.023 \\
    5 & 0.010 & 0.011\\
    6 & 0.006 & 0.007\\
    7 & 0.003 & 0.004\\
    8 & 0.002 & 0.003\\
    9 & 0.001 & 0.002\\
    >=10 & 0.009 & 0.020\\
    }\newentdata

\begin{tikzpicture}[scale=0.7]
    \begin{axis}[
            ybar,
            bar width=0.2 cm,
            legend style={at={(0.6,0.99)},
                anchor=north},
            every tick/.style={
        	black,
        	semithick,
      		},
            symbolic x coords={0, 1, 2, 3, 4, 5, 6, 7, 8, 9, >=10},
            xtick=data,
            xticklabel style={rotate=30, anchor=east, align=right},
            ylabel={$\text{Probability}$ },
            xlabel = No of mentions,
        ]
        \addplot table[x=rng,y=acc]{\newentdata};
        \addplot table[x=rng,y=tar]{\newentdata};
        \legend{Account Holder, Target}
    \end{axis}
\end{tikzpicture}

\caption{Comparison between number of mentions in a tweet by the account holder vs the target accounts.}
\label{fig:Mentionscdf}
    \end{minipage}%
\end{figure*}

\vspace{-3mm}
\subsection{Tweeting behavior}
The following analysis has been done on the timeline crawls of the account holders and the targets to identify differences in their respective tweeting patterns. Figure~\ref{fig:Statuscdf} shows a comparison between the account holder and the target accounts with respect to their tweeting behavior. We observe that there are more account holders who tweet less or moderately compared to the targets ($p$-value of significance is $<10^{-5}$). However, in the high tweeting zone, there are comparatively more targets than the account holders. In fact, there are several accounts with more than 100K tweets. These are typically news media accounts like Fox News, Sky Sports, Telegraph, CBS Sweden, CNN News 18 etc. They are being cyberaggressed for the news/opinions the news anchors associated with them typically express. We cite below such an example of a news media involved in incivility.
\begin{tcolorbox}[boxsep=0pt, top=1pt,left=1pt,right=1pt,bottom=1pt,colframe=gray!50]\footnotesize{
\noindent @user1 @newsmedia1 stupid English b*tch asking is there people on that plane?... No you thick c*nt it's like google cars....\\
\textbf{Incivility context:}\\
A Dubai firefighter has died of injuries sustained putting out fire after plane crash landing - Emirates chairman https://t.co/i25sjAKfOC}
\end{tcolorbox}

We also observe the mention patterns of the targets and account holders.
Figure~\ref{fig:Mentionscdf} shows the distribution of number of mentions in a tweet by the account holders versus the distribution of number of mentions in a tweet by the targets. It is evident that there are many targets ($\sim30\%$) who do not mention at all. The account holders generally mention multiple people in a tweet and much more often than the targets ($p$-value of significance is $<10^{-5}$). This also indicates that they might be insulting multiple targets. We further observe that there are some targets who use more mentions in their tweets than account holders. These targets have high followers (198K on average) and they probably use more mentions mostly for promotional purposes.
\subsection{Linguistic and cognitive dimensions} 
\begin{figure*}[t] 
\centering
\subcaptionbox{Comparison of LIWC scores in the linguistic categories between the account holder and the target tweets.\label{fig:ling}}%
  [0.3\textwidth]
  {
  	\pgfplotstableread[row sep=\\,col sep=&]{
    rng & acc & tar & acc_err & tar_err\\
    1st person singular & 0.023 & 0.022 & 0.0003 & 0.0003\\
    1st person plural & 0.006 & 0.008 & 0.0001 & 0.0002\\
    2nd person & 0.021 & 0.016 & 0.0003 & 0.0003\\
    3rd person singular & 0.009 & 0.007 & 0.0002 & 0.0002\\
    3rd person plural & 0.005 & 0.004 & 0.0001 & 0.0001\\
    Negations & 0.015 & 0.012 & 0.0003 & 0.0002\\
    Swear Words & 0.004 & 0.002 &  0.0002 & 0.0002\\
    }\newentdata

\begin{tikzpicture}[scale=0.55]
    \begin{axis}[
            ybar,
            bar width=0.2 cm,
            legend style={at={(0.7,0.99)},
                anchor=north},
            every tick/.style={
        	black,
        	semithick,
      		},
            symbolic x coords={1st person singular, 1st person plural, 2nd person, 3rd person singular, 3rd person plural, Negations, Swear Words},
            xtick=data,
            xticklabel style={rotate=90, anchor=east, align=right, text width=8em},
            ylabel={$\text{LIWC Values}$ },
        ]
        \addplot+[error bars/.cd, error mark=-, y dir=both, y explicit] table[x=rng,y=acc, y error=acc_err]{\newentdata};
        \addplot+[error bars/.cd, error mark=-, y dir=both, y explicit] table[x=rng,y=tar, y error=tar_err]{\newentdata};
        \legend{Account Holder, Target}
    \end{axis}
\end{tikzpicture}
}  
  \hfill
\subcaptionbox{Comparison of LIWC scores in personal concerns categories between account holder and target tweets.\label{fig:person}}
  [0.3\textwidth]{
    {
  	\pgfplotstableread[row sep=\\,col sep=&]{
    rng & acc & tar & acc_err & tar_err\\
    Work & 0.007 & 0.008 & 0.0001 & 0.0002\\
    Achieve & 0.008 & 0.009 & 0.0002 & 0.0002\\
    Leisure & 0.009 & 0.008 & 0.0002 & 0.0002\\
    Home & 0.001 & 0.001 & 0.00008 & 0.00009\\
    Money & 0.005 & 0.004 & 0.0001 & 0.0001\\
    Religion & 0.002 & 0.002 & 0.0001 & 0.0001\\
    Death & 0.002 & 0.001 &  0.00008 & 0.00008\\
    }\newentdata

\begin{tikzpicture}[scale=0.55]
    \begin{axis}[
            ybar,
            bar width=0.2 cm,
            legend style={at={(0.7,0.99)},
                anchor=north},
            every tick/.style={
        	black,
        	semithick,
      		},
            symbolic x coords={Work, Achieve, Leisure, Home, Money, Religion, Death},
            xtick=data,
            xticklabel style={rotate=90, anchor=east, align=right, text width=8em},
            ylabel={$\text{LIWC Values}$ },
        ]
        \addplot+[error bars/.cd, error mark=-, y dir=both, y explicit] table[x=rng,y=acc, y error=acc_err]{\newentdata};
        \addplot+[error bars/.cd, error mark=-, y dir=both, y explicit] table[x=rng,y=tar, y error=tar_err]{\newentdata};
        \legend{Account Holder, Target}
    \end{axis}
\end{tikzpicture}
}  
  }
  \hfill
  \subcaptionbox{Comparison of LIWC scores in the cognitive categories between the account holder and the target tweets.\label{fig:psych}}%
  [0.35\textwidth]
  {
  	\pgfplotstableread[row sep=\\,col sep=&]{
    rng & acc & tar & acc_err & tar_err\\
    Social & 0.070 & 0.062 & 0.0006 & 0.0006\\
    Family & 0.002 & 0.002 & 0.0001 & 0.0001\\
    Friend & 0.0005 & 0.0007 & 0.00006 & 0.00008\\
    Humans & 0.006 & 0.006 & 0.0002 & 0.0002\\
    Positive Emotions & 0.031 & 0.033 & 0.0006 & 0.0006\\
    Negative Emotions & 0.014 & 0.009 & 0.0003 & 0.0003\\
    Anxiety & 0.0007 & 0.0006 &  0.00007 & 0.00007\\
    Anger & 0.007 & 0.004 &  0.0003 & 0.0002\\
    Sadness & 0.012 & 0.012 &  0.0003 & 0.0003\\
    Perceptual & 0.005 & 0.006 &  0.0002 & 0.0002\\
    See & 0.003 & 0.003 &  0.0001 & 0.0001\\
    Hear & 0.003 & 0.003 &  0.0001 & 0.0001\\
    Feel & 0.011 & 0.009 &  0.0003 & 0.0003\\
    Biological & 0.004 & 0.003 &  0.0002 & 0.0002\\
    Body & 0.002 & 0.001 &  0.0001 & 0.0001\\
    Health & 0.005 & 0.003 &  0.0002 & 0.0002\\
    }\newentdata

\begin{tikzpicture}[scale=0.55]
    \begin{axis}[
            ybar,
            bar width=0.05 cm,
            legend style={at={(0.7,0.99)},
                anchor=north},
            symbolic x coords={Social, Family, Friend, Humans, Positive Emotions, Negative Emotions, Anxiety, Anger, Sadness, Perceptual, See, Hear, Feel, Biological, Body, Health},
            xtick=data,
            xticklabel style={rotate=90, anchor=east, align=right, text width=8em},
            ylabel={$\text{LIWC Values}$ },
            every tick/.style={
        black,
        semithick,
      },
        ]
        \addplot+[error bars/.cd, error mark=-, y dir=both, y explicit] table[x=rng,y=acc, y error=acc_err]{\newentdata};
        \addplot+[error bars/.cd, error mark=-, y dir=both, y explicit] table[x=rng,y=tar, y error=tar_err]{\newentdata};
        \legend{Account Holder, Target}
    \end{axis}
\end{tikzpicture}
}  
  \vspace{-3mm}\caption{LIWC analysis. We have used standard error as error bars in the graph.}
    \label{fig:lingpsych}
    \vspace{-0.2cm}
\end{figure*}
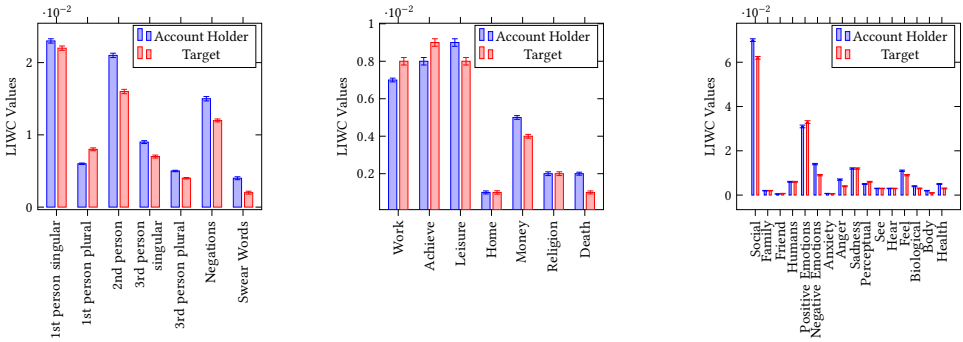

In this section, we perform linguistic and psychological measurements of the tweets posted by the target and the account holder profiles. We use the Linguistic Inquiry and Word Count (LIWC)~\cite{pennebaker2001linguistic}, a text analysis software to find which categories of words are used by the targets and the account holders. LIWC evaluates different aspects of word usage in different meaningful categories, by counting the number of words across the text for each category. Table~\ref{tab:liwccat} notes the frequent words from our dataset corresponding to the different LIWC categories. In Figure~\ref{fig:ling}, we show the linguistic categories for the target and the account holder tweets. It is quite evident from the figure that the account holders use first person quite frequently compared to the targets and refer to others (here, targets) in third person. Moreover, the use of swear words and negations are more in the account holder tweets than the target tweets.


Next, we look at some of the personal concerns such as work, achievements, leisure, etc. Figure~\ref{fig:person} shows the differences of the account holder and the target profiles in terms of expression of personal concerns. We observe that the targets tweet more about work and achievements; account holders tweet more about  monetary aspects. We also observe that in ``religion'' (e.g., church, mosque) and ``death'' (e.g., bury, kill) categories, account holders tweet more than the targets. This is indicative of the fact that religion-based incivility is more prevalent in social media. This is in line with the work by Hosseinmardi et al.~\shortcite{hosse_religion} where they found that there is a high usage of profane words around words like ``muslim'' in ask.fm social media.

We further analyze the cognitive aspects of the target and the account holder tweets (see Figure~\ref{fig:psych}). We observe that account holders are more expressive of their emotions than the targets. Further, the account holders tend to tweet more related to ``body'' (e.g. face, wear) and ``sexual'' (e.g. sl*t, rapist etc.) categories. In ``friend'' category, both account holder and target tend to tweet in similar proportions. However, the account holders talk more frequently related to the ``social'' category. Also targets express more ''positive emotion'' whereas, account holders express ''negative emotion''.
\begin{table}[h]
 \centering
 \caption{\label{tab:liwccat}Frequent words from the LIWC category in our dataset. The asterisk here denotes the acceptance of all letters, hyphens, or numbers following its appearance.\protect\footnotemark}
\tiny
 \begin{tabular}{|c|c|} 
 \hline
Category & Frequent words from the dataset \\\hline
Swear words & shit*, dumb*, bloody, crap, fuck \\
Work& Read, police, political, policy, student* \\
Achieve& Better, win, won, first, best \\
Leisure& party*, read, running, show, shows \\
Home & clean*, address, home, family, house* \\
Money & Free, money*, worth, trade*, tax \\
Religion& Sacred, moral, worship*, hell, devil* \\
Death& War, death*, murder*, kill*, die \\
Social&You, we, your, our, they \\
Family&Family, families*, pa, mother, ma \\
Friends& Mate, mates, fellow*, lover*, friend* \\
Humans& people*, human*, women*, children*, woman \\
Positive Emotions& Like, party*, lol, better, support \\
Negative Emotions& War, wrong*, violent*, liar*, rape* \\
Anxiety& doubt*, fear, risk*,avoid*, afraid \\
Anger& War, violent*, liar*, rape*, fight* \\
Sadness&low*, lost, lose, loser*, fail*\\
Perceptual&green*, say*, said, watch*, see \\
See&green*, watch*, see, white*, look \\
Hear&say*, said, hear, heard, listen \\
Feel&round*, hard, loose*, hand, feel \\
Biological & health*, drug*, rape*, life, shit* \\
Body & shit*, head, brain*, hand, face \\
Health &health*, drug*, life, weak*, living \\ \hline
\end{tabular}
\end{table}


\section{Detecting incivility}
In the earlier section, we have discussed the possible connections of incivility with target dependent sentiments and observed the various behavioral aspects of the targets and the account holders. Note that to do this, we had to manually label the incivil tweets as a first step in order to build the incivility context. In this section, we shall propose an automatic approach to classify incivil tweets.

\noindent \textbf{Training and test sets:}
Recall that we have 24,271 manually labeled tweets. We choose randomly 21,000 of these tweets and consider them as our training set. The remaining 3271 points are in the test set. 


\subsection{Baseline models}
There have been several studies like~\cite{cbullyingml} that consider identifying incivility instances in various social media platforms based on various content features. We adopt this work (baseline 1) for comparison with our model. A few studies use basic n-gram features. Hosseinmardi et al.~\shortcite{hosseinmardi2015detection} uses unigrams and tri-grams to detect incivility. Xu et al.~\shortcite{xu2012learning} uses unigrams, unigrams+bigrams as features to detect incivil traces in social media. These n-gram features constitute our second baseline model. We also consider n-gram (uni-, bi-,tri-) based model with automatic feature selection to reduce the feature space to smaller number of features.
We also use the work by \citet{chen2017presenting} as a baseline. In particular, we use n-grams together with textual features. The textual features include number of words in the tweet, number of characters in the tweet, 
number of sentences in the comment
,average word length (\#characters divided by \#words)
, average sentence length (\#words divided by \#sentences)
, profane words usage level (\#profane words divided by \#words)
, uppercase letter usage (\#uppercase letters divided by \#sentences)
, punctuations usage (\#punctuations divided by \#sentences)
, URL usage level (\#URL divided by \#words)
, mentions usage level (\#mentions divided by \#words)
For all the above models, we use different classifiers like SVM with linear and rbf kernels, $k$-NN, logistic regression, Adaboost with L2 regularization wherever possible and report the best overall accuracy.

\noindent\textit{Feature engineering}: We use the following content features adopted from Bommerson~\shortcite{cbullyingml}.\\
\noindent\textbf{Length of the tweet.} This feature simply corresponds to the length of the tweet measured in terms of the number of words in the tweet.

\noindent\textbf{Number of negative/offensive words in the tweet.} This feature calculates the number of offensive words using the offensive words list we introduced earlier.

\noindent\textbf{Severity value of the tweet.} We label every word in the offensive words list manually as `1' or `2' and this is unanimously agreed upon by two of the authors. This is done because all offensive words are not
equally bad; some have more severe effect when used in the tweet. We label the more severe offensive words as `2' while the less severe ones are labeled as `1'. 
If $w$ is a bad word in a tweet and $W$ is a set of all offensive words, then 
severity of a tweet = $\sum\limits_{w \in W} sev(w)$, where $sev(w)$ denotes the severity of the word $w$ and $sev(w) \in \{1, 2\}$.

\noindent \textbf{Time of the tweet.} The time of post of a tweet is taken as a feature. It can be a distinguishing feature between the two classes as incivility among teenagers happens usually after school whereas normal tweets flow in all throughout the day.

\noindent \textbf{Negation used or not.} Sometimes the offensive words can be used with negations. This however reverses the effect of the word and thus the tweets in these cases are usually civil (e.g., `@xyz Please don't die.').  To handle this case, we take into account if negations are used along with offensive words.

\noindent\textit{Drawbacks of the baseline models}:
The baseline model suffers from the following drawbacks:


\noindent \textbf{Sentence dependencies.} The above features cannot detect dependencies between different sentences in a tweet. For example, `@xyz you are a footballer. All footballers are brainless.'

\noindent \textbf{Multiple connotations.} The above features do not take into account the multiple connotations of a word or phrase. For example, the word `well' has a neutral connotation whereas the words `sooo welllll' indicates a much happier/excited state of the mind of the author. 

\noindent \textbf{Obfuscated words.} The account holder can use words like `a\$\$hole' or `f**k' which will not get detected by the above features.

In this work, we use character-level models that build representations of sentences using the constituent characters. The advantage of using these kinds of models is that we can easily deal with non-traditional spellings used in social media website and hence, can easily capture obfuscated words, multiple connotations of words and sentence dependencies without the need of normalizing the words to their correct forms. In particular, we use several deep learning models that take such character embeddings as input and output a sentence representation using which we are able to classify if the tweet is related to incivility or not. As an additional deep learning based basline we use the model proposed by Pavlopoulos et al.~\cite{pavlopoulos2017deeper} that was introduced to moderate abusive user content.

\subsection{The deep learning frameworks}
We propose deep learning frameworks primarily based on character-level LSTM and character-level CNNs. We use character-level models in this experiment as the characters can provide more information than the words as a whole for the tweet dataset under consideration. In particular, since the tweets are limited to only 140 (now 280) characters, it is very difficult to obtain rich embeddings from such a short text where only a few words are present. Therefore, character-level LSTM/CNN can obtain more information than a word-level LSTM/CNN in such a scenario. We would also like the model to have knowledge about the past sequence of characters as well as to look into the future. Thus, we choose the bidirectional LSTM framework\footnote{We have also compared the model with unidirectional LSTM, GRUs although we do not report the results for all these.} for our purpose. 

The second model we use is a character-level CNN model. The authors in~\cite{zhang2015character} have successfully used character level CNNs for text classification. This motivated us to try CNNs for the purpose of our task. We use a model with 2 convolution layers, each layer followed by a max-pooling layer. 

\noindent\textbf{Bi-directional LSTM}: We use a character-level bi-directional LSTM with {\sl tanh} as the function that provides non-linearity. The 100 dimensional characters embeddings are initialized randomly and updated during training.  The encoded vectors are passed to the LSTM units and then the output of each unit which is a 100 dimensional vector is passed to the next unit in the forward as well as backward direction. We concatenate the output of the LSTM unit in the last timestep in both forward and backward direction.  Finally, we use  a softmax layer to calculate the probability of the tweet belonging to a particular class. 

\noindent\textbf{Character CNN}:
The second model we use is a character-level CNN model with ReLU non-linearity. The first convolution layer contains 100 filters with kernel size $5\times5$. The second convolution layer, once again, contains 100 filters with kernel size $5\times5$. Both the convolution layers are followed by a max pooling layer. Finally, we flatten the output vector and use a dense layer followed by a softmax to get the output probabilities for the two classes. We also use dropout and batch normalization in both the above models to prevent overfitting.

\begin{figure}[t]
    \centering
\includegraphics[scale=0.35,angle=-90]{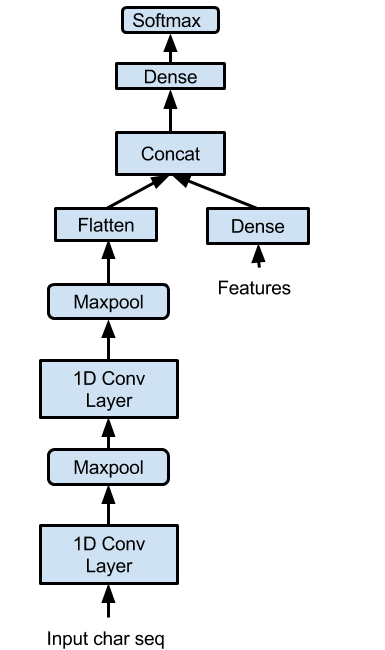}
\caption{char-CNN + opinion conflict feature based model for incivility detection.}
\label{fig:arch}
\end{figure}

\noindent\textbf{Character CNN + opinion conflict feature}: The char-LSTM and char-CNN models might be able to extract some features and temporal relations between characters but still they might not be able to take into account the crucial observations about incivility that we made in the two previous sections. To improve the performance of the model and to take advantage of both the scenarios, in the char-CNN model, we fuse the flattened vector we get after the two convolution and maxpool layers with a special entity sentiment based feature, which we have seen in the previous sections to be very discriminative. The full architecture is shown in Figure~\ref{fig:arch}. In particular, we take the vector obtained by flattening and concatenate it with the \textit{opinion conflict feature} described below. Then we take this vector as an input to a feed forward neural network which classifies the tweet as an instance of incivility or otherwise. Thus, we train the fused model end-to-end according to the final loss function. The weights learned from the feed forward network at the end of the concatenated flattened vector and custom feature vector tell the model the amount by which each feature should be weighed.  

\noindent\textit{Opinion conflict feature}: The opinion conflict feature is the total count of opinion conflicts that appear in the incivility context between the account holder and all the targets in a particular tweet.

\noindent\textbf{Training: }  We perform all parameter updates through backpropagation and \emph{stochastic gradient descent}. We apply dropout with probability $0.25$. We also use early stopping with patience value as 3.

\begin{table}[h]
 \centering
 \caption{Classification results}~\label{tab:table1}
\resizebox{14cm}{!}{
\begin{threeparttable} 
 \begin{tabular}{|c|c|c|c|} 
 
 \hline
   Method & Accuracy & F1-Score & ROC-AUC\\
 \hline
   \multicolumn{4}{|c|}{Content based features}\\
   \hline
   \cellcolor{pink} Bommerson et al. ~\shortcite{cbullyingml}) & \cellcolor{pink}71.1\% & \cellcolor{pink}0.27 & \cellcolor{pink}0.61\\
   \hline
   \hline
   \multicolumn{4}{|c|}{n-gram based features}\\
   \hline
   unigrams & 86.4\% & 0.73 & 0.91\\
   unigrams (automatic feature selection) & 85.6\% & 0.72 & 0.91\\
   bigrams & 86.7\% & 0.66 & 0.79\\
   bigrams (automatic feature selection) & 88.4\% & 0.69 & 0.81\\
   trigrams & 82.3\% & 0.18 & 0.54\\
   trigrams (automatic feature selection) & 83.6\% & 0.29 & 0.59\\
   unigrams + bigrams & 86.7\% & 0.74 & 0.91\\
   unigrams + bigrams (automatic feature selection) (\citet{xu2012learning}) & 87.5\% & 0.67 & 0.79\\
   unigrams + trigrams & 86.7\% & 0.63 & 0.62\\
   unigrams + trigrams (automatic feature selection) (\citet{hosseinmardi2015detection}) & 86.7\% & 0.64 & 0.62\\
   bigrams + trigrams & 86.8\% & 0.66 & 0.79 \\
   bigrams + trigrams (automatic feature selection) & 88.8\% & 0.76 & 0.93 \\
   unigrams + bigrams + trigrams & 86.8\% & 0.73 & 0.91\\
   \cellcolor{yellow}unigrams + bigrams + trigrams (automatic feature selection) & \cellcolor{yellow}88.9\% & \cellcolor{yellow}0.77 & \cellcolor{yellow}0.92\\
     baseline 1 + opinion conflict & 76.1\% & 0.37 & 0.61\\
     unigrams + bigrams + trigrams + opinion conflict & 80.9\% & 0.77 & 0.90\\
     unigrams + bigrams + trigrams + textual features (\citet{chen2017presenting}) & 86.7\% & 0.77 & 0.92\\

   \hline \hline
   \multicolumn{4}{|c|}{char-LSTM and char-CNN Models \tnote{1}}\\
   \hline
char-LSTM & 88.9\% & 0.80  & 0.84\\
char-LSTM+attention (\citet{pavlopoulos2017deeper}) & 78.5\% & 0.53 & 0.74\\
char-LSTM+attention (\citet{pavlopoulos2017deeper})+opinion conflict & 78.6\% & 0.54 & 0.75\\
char-CNN & 93.0\% & 0.81 & 0.88\\
\cellcolor{green}char-CNN + opinion conflict & \cellcolor{green}\textcolor{black}{\bf 93.3\%} & \cellcolor{green}\textcolor{black}{\bf 0.82} & \cellcolor{green}\textcolor{black}{\bf 0.89}\\
 \hline
\end{tabular}
\begin{tablenotes}
\item[1] Results have been obtained by taking mean of 10 random runs. 
\end{tablenotes}
\end{threeparttable}
}
\vspace{-3mm}

\end{table}
\noindent\textbf{Results}: 
We train the model in a supervised learning framework using the same training and test data on which we had trained the baseline model. The results are noted in Table~\ref{tab:table1}. The first observation is that content based features do not perform very well in this setting (see \textcolor{pink}{pink} row).

As we can see from the table, the character CNN model along with the opinion conflict feature (see {green} row) outperforms the baseline models achieving {93.3\%} accuracy with F1-score of {0.82} and ROC area of {0.89}. Note that while the increase in F1-score from char-CNN to char-CNN + opinion conflict model is from 0.81 to 0.82, it is statistically significant with $p$-value $<0.05$. Furthermore the char-CNN + opinion conflict model corrects 10.14\% of the errors made by the char-CNN model.

Among the baseline models, the best performance is obtained when we use unigram, bigram and trigrams as features (see {yellow} row). Our model significantly outperforms this best performing baseline ({5.1\%}, {6.5\%} improvements w.r.t accuracy and F1-score). The CNN model seems to learn better representations than hand picked bag of word n-gram features. 

Below we list a couple of examples where char-CNN fails and char-CNN+opinion conflict predicts correctly. Both cases have very few linguistic cues to help in the classification task. In the first case, the account holder tweets negatively about Old Alabama whereas the target @user1 tweets positively about Old Alabama. In the second case, we found that the account holder tweets positively about Daily News and @user1, the target negatively about Daily News.

\begin{tcolorbox}[boxsep=0pt, top=1pt,left=1pt,right=1pt,bottom=1pt,colframe=gray!50]\footnotesize{
1. @user1 @user2 YOU and your incessant attempts at lying to and bullying the press while at the White House as PS are reason for apology you sack of dog stools. The AMERICAN public does not distrust the media  other than by your disingenuous propaganda. Ur time is coming you asshole. \\ 
\smallskip
2. @user1 No just you. Everyone would know he was asking for you. Asshole!
}
\end{tcolorbox}

\vspace{-3mm}
\section{Post-hoc analysis} 
We run the trained char-CNN model on the entire dataset of $\sim300,000$ tweets and perform a post-hoc analysis of the incivility tweets to extract some more interesting properties. The key observation is that there are multiple instances of incivil posts from the same account holder as well as a target being attacked multiple times. 

\vspace{-3mm}
\subsection{Repetition of incivility and targeting}
In this section, we analyze the incivil posts in more detail focusing on their repetitions. Following are some examples of multiple incivility and targeting. The first two examples show @user1 being attacked by multiple account holders (user3 and user4) and the last two examples show @user5 being attacked multiple times by user6.
\begin{figure}[!htbp]
    \centering
    \begin{minipage}[t]{0.47\textwidth}
\pgfplotstableread[row sep=\\,col sep=&]{
    rng & acc  \\
    2 & 0.69  \\
    3 & 0.17  \\
    4 & 0.07\\
    5 & 0.03 \\
    6 & 0.02 \\
    7 & 0.009 \\
    8 & 0.006\\
    9 & 0.004 \\
    10 & 0.003 \\
    11 & 0.002 \\
    }\newentdata

\begin{tikzpicture}[scale=0.7]
    \begin{axis}[
            ybar,
            bar width=0.2 cm,
            legend style={at={(0.18,0.99)},
                anchor=north},
            every tick/.style={
        	black,
        	semithick,
      		},
            symbolic x coords={2, 3, 4, 5, 6, 7, 8, 9, 10, 11},
            xtick=data,
            xticklabel style={rotate=0, anchor=east, align=right},
            ylabel={$\text{Fraction of Account Holders}$ },
            xlabel= No of times an account holder harasses,
        ]
        \addplot+[color=blue, fill=blue] table[x=rng,y=acc]{\newentdata};
    \end{axis}
\end{tikzpicture}

\caption{Account holders involved in multiple instances of incivility.
}
\label{fig:mulbully}
\end{minipage}
\hfill \hfill
\begin{minipage}[t]{0.47\textwidth}
\centering
\pgfplotstableread[row sep=\\,col sep=&]{
    rng & acc  \\
    2 & 0.54  \\
    3 & 0.19  \\
    4 & 0.09\\
    5 & 0.06 \\
    6 & 0.04 \\
    7 & 0.02 \\
    8 & 0.02\\
    9 & 0.01 \\
    10 & 0.01 \\
    11 & 0.01 \\
    }\newentdata

\begin{tikzpicture}[scale=0.7]
    \begin{axis}[
            ybar,
            bar width=0.2 cm,
            legend style={at={(0.18,0.99)},
                anchor=north},
            every tick/.style={
        	black,
        	semithick,
      		},
            symbolic x coords={2, 3, 4, 5, 6, 7, 8, 9, 10, 11},
            xtick=data,
            xticklabel style={rotate=0, anchor=east, align=right},
            ylabel={$\text{Fraction of Targets}$ },
            xlabel= No of times a target gets harassed,
        ]
        \addplot+[color=red, fill=red] table[x=rng,y=acc]{\newentdata};
    \end{axis}
\end{tikzpicture}
\caption{Targets attacked multiple times.}
\label{fig:multarget}
\end{minipage}
\vspace{-5mm}
\end{figure}

\begin{tcolorbox}[boxsep=0pt, top=1pt,left=1pt,right=1pt,bottom=1pt,colframe=gray!50]\footnotesize{
\noindent @user1 Hillary was a skankb*tch,and they make a cream for your butt-hurt condition (account holder: user3)\\
\noindent @user1 Get ur a\$\$ off here, u stupid b*tch (account holder: user4) \\
\\
\noindent @user5 you are the biggest f**king piece of scum there is. Karma is a b*tch (account holder: user6)\\ 
\noindent @user5 you are the biggest scumbag (account holder: user6)}
\end{tcolorbox}
In Figure~\ref{fig:mulbully} we show the distribution of account holders involved in multiple instances of attacking another person. We observe that there are a significant number of account holders ($\sim13\%$) who attack more than once and if we observe the distribution among them, while a majority of them attack 2-5 times, there are also account holders who attack more than 10 times. This scenario of multiple incivil posts seems alarming as it might be one of the prime reason for users to quit social media sites. Thus, it is important to detect these account holders' accounts and take necessary preventive measures.

We also observe that the same target is attacked multiple times. In Figure~\ref{fig:multarget}, we see that there are a significant number of targets ($\sim19\%$) who get attacked multiple times and a majority of them experience 2-3 times of attack. However, there are also a small fraction of targets who are attacked more than 10 times.

\subsection{Reputation scores}
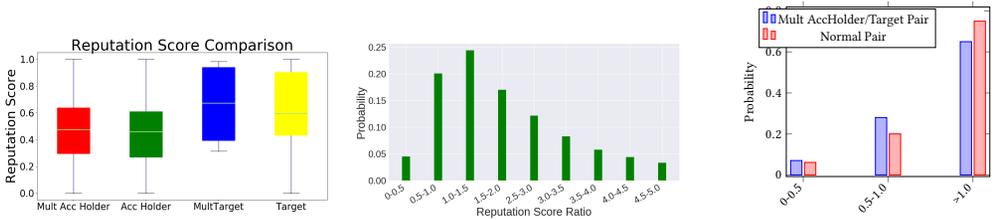
\begin{figure*}[t] 
\centering
\subcaptionbox{Reputation score comparison between account holders, targets, account holders offending multiple-time times and multiple-time targets.
\label{fig:multrepscore}}%
  [0.32\textwidth]{\includegraphics[scale = 0.14]{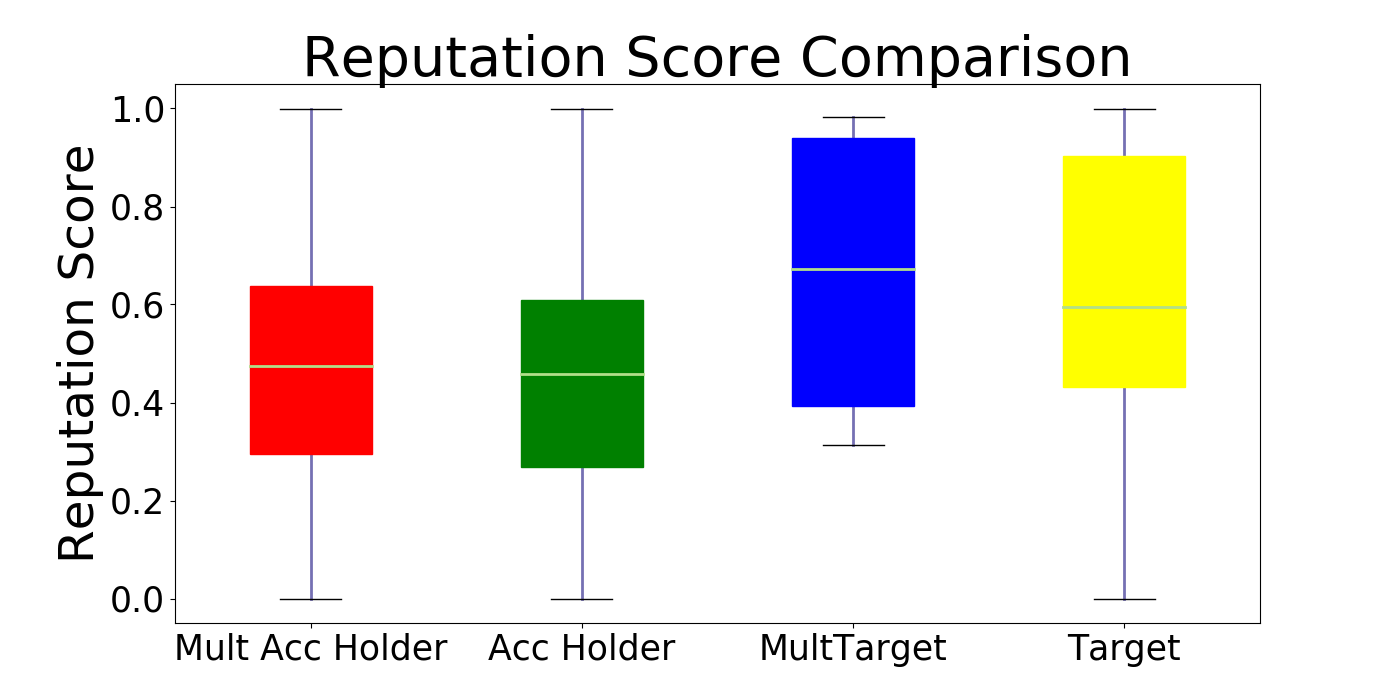}}
  \hfill
\subcaptionbox{Reputation score comparison for account holder-target pairs.\label{fig:pair}}
  [0.32\textwidth]{\includegraphics[scale = 0.14]{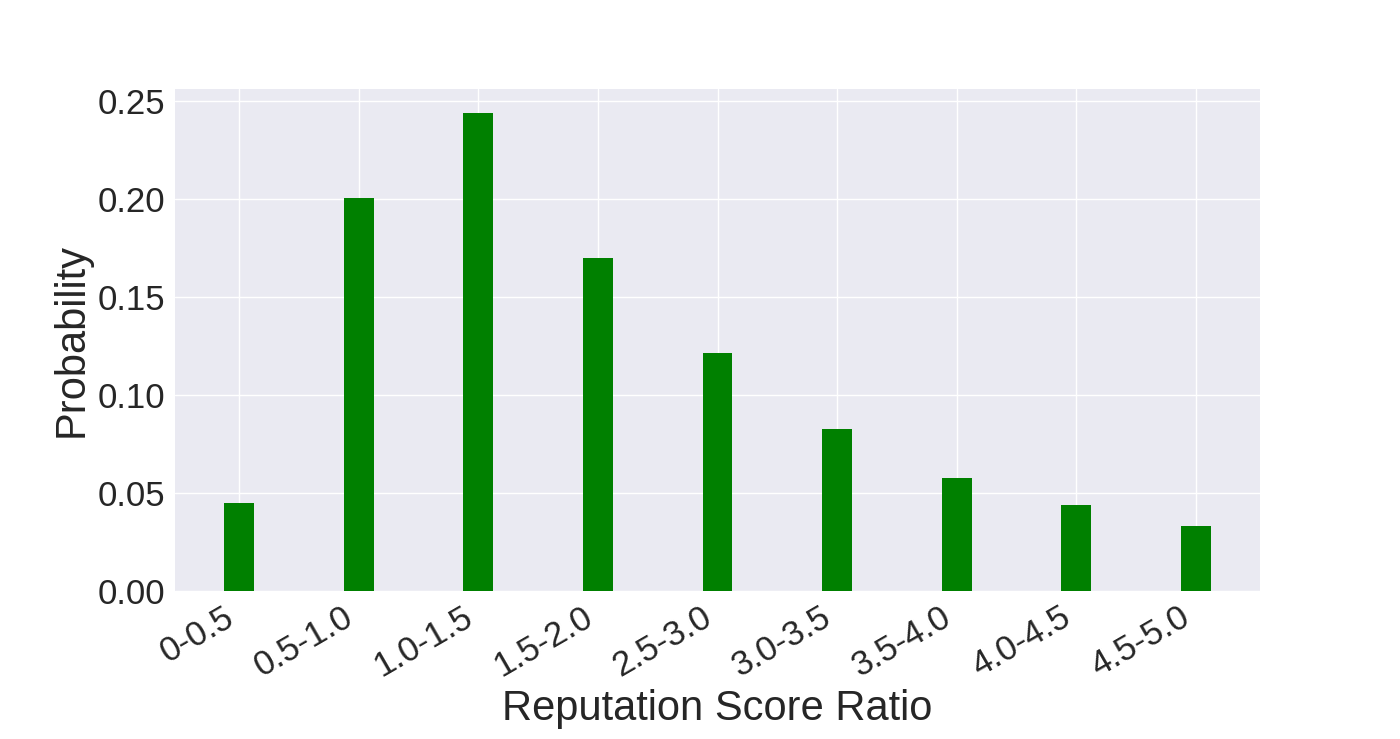}}
  \hfill
  \subcaptionbox{Reputation score comparison for account holder-target pairs appearing multiple times and pairs appearing once.\label{fig:repscore}}%
  [0.32\textwidth]
 {
 \pgfplotstableread[row sep=\\,col sep=&]{
    NE & neu & pos \\
    0-0.5     & 0.07 & 0.061 \\
    0.5-1.0 & 0.28 & 0.2 \\
    >1.0     & 0.65 & 0.75\\
    }\newentdata

\begin{tikzpicture}[scale=0.5]
    \begin{axis}[
            ybar,
            bar width=0.3 cm,
            legend style={at={(0.3,0.99)},
                anchor=north},
            symbolic x coords={0-0.5, 0.5-1.0, >1.0},
            xtick=data,
            xticklabel style={rotate=45, anchor=east},
            ylabel={$\text{Probability}$ },
        ]
        \addplot table[x=NE,y=neu]{\newentdata};
        \addplot table[x=NE,y=pos]{\newentdata};
        \legend{Mult AccHolder/Target Pair, Normal Pair}
    \end{axis}
\end{tikzpicture}
 }
  \vspace{-3mm}\caption{Reputation score comparison}
    \label{fig:rscore}
    \vspace{-0.2cm}
\end{figure*}

Reputation score is a measure that determines the reputation of a profile in social media and is defined as $\frac{\#followers}{\#followers + \#friends}$. A celebrity who tends to have a very large followership but follows less people, i.e., having $followers$ $>>$ $friends$ will have a reputation score close to 1. Figure~\ref{fig:multrepscore} shows the average reputation scores for the account holders, the targets, the multiple-time insulting account holders and the multiple-time targets. The plot suggests that targets who are attacked multiple times have the highest reputation scores and are therefore celebrities with high probability. Moreover, the targets in general tend to have higher reputation scores than the account holders. We next consider the account holder-target pairs. Figure~\ref{fig:pair} shows the distribution of the reputation score ratio ($\frac{rep_{score}(target)}{rep_{score}(account holder)}$) for the account holder-target pairs. It is quite evident from the result that there are more number of account holder-target pairs with reputation score ratio more than 1. Therefore, in general, the targets are more reputed than account holders. Figure~\ref{fig:repscore} shows the distribution of reputation score ratios for the account holder-target pairs appearing multiple times and pairs appearing once. The pairs occurring multiple times tend to have greater reputation score ratio with higher probability. Thus, targets who are attacked by the same account holder again and again seem to have higher number of followers, and are highly likely to be celebrities.

\subsection{Follower/followee properties}
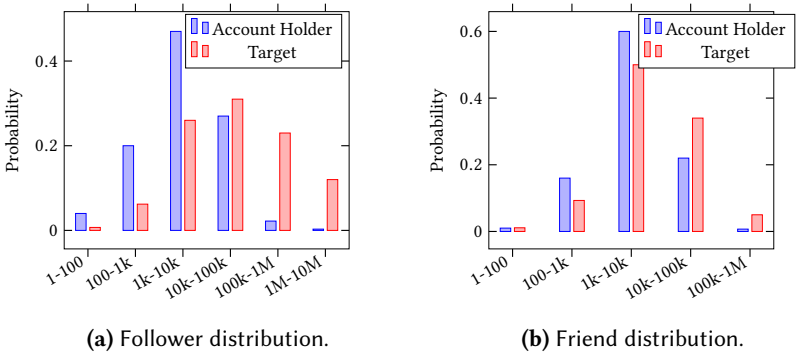
\begin{figure}[h] 
\centering
\begin{subfigure}{0.4\textwidth}
\pgfplotstableread[row sep=\\,col sep=&]{
    rng & acc & tar \\
    1-100     & 0.04 & 0.007 \\
    100-1k & 0.20 & 0.062 \\
    1k-10k     & 0.47 & 0.26 \\
    10k-100k & 0.27 & 0.31 \\
    100k-1M & 0.022 & 0.23 \\
    1M-10M & 0.003 & 0.12\\
    }\newentdata

\begin{tikzpicture}[scale=0.7]
    \begin{axis}[
            ybar,
            bar width=0.2 cm,
            legend style={at={(0.7,0.99)},
                anchor=north},
            every tick/.style={
        	black,
        	semithick,
      		},
            symbolic x coords={1-100, 100-1k, 1k-10k, 10k-100k, 100k-1M, 1M-10M, 10M-100M},
            xtick=data,
            xticklabel style={rotate=30, anchor=east, align=right},
            ylabel={$\text{Probability}$ },
        ]
        \addplot table[x=rng,y=acc]{\newentdata};
        \addplot table[x=rng,y=tar]{\newentdata};
        \legend{Account Holder, Target}
    \end{axis}
\end{tikzpicture}
\caption{Follower distribution.}
\label{fig:multfollow}
\end{subfigure}
\begin{subfigure}{0.4\textwidth}
\pgfplotstableread[row sep=\\,col sep=&]{
    rng & acc & tar \\
    1-100     & 0.01 & 0.011 \\
    100-1k & 0.16 & 0.093 \\
    1k-10k     & 0.60 & 0.50 \\
    10k-100k & 0.22 & 0.34 \\
    100k-1M & 0.007 & 0.05 \\
    }\newentdata

\begin{tikzpicture}[scale=0.7]
    \begin{axis}[
            ybar,
            bar width=0.2 cm,
            legend style={at={(0.8,0.99)},
                anchor=north},
            every tick/.style={
        	black,
        	semithick,
      		},
            symbolic x coords={1-100, 100-1k, 1k-10k, 10k-100k, 100k-1M, 1M-10M, 10M-100M},
            xtick=data,
            xticklabel style={rotate=30, anchor=east, align=right},
            ylabel={$\text{Probability}$ },
        ]
        \addplot table[x=rng,y=acc]{\newentdata};
        \addplot table[x=rng,y=tar]{\newentdata};
        \legend{Account Holder, Target}
    \end{axis}
\end{tikzpicture}
\caption{Friend distribution.}
\label{fig:multfriend}
\end{subfigure}
\vspace{-3mm}\caption{Distribution of various properties for multiple-time account holders and multiple-time targets.}
    \label{fig:multbullvict}
    \vspace{-0.2cm}
\end{figure}

We further study the follower/followee behavior of multiple-time insulting account holders and targets. Figure~\ref{fig:multfollow} shows a comparison between the followers of account holders who attack multiple times and targets who are targeted multiple times. The graph shows that the targets have higher number of followers in general as compared to the account holders ($p$-value of significance is $<10^{-5}$). Many of the targets' follower count falls in the 1M -- 10M range, whereas more than 40\% of the account holders' follower count lies in the 1K-10K bucket. Figure~\ref{fig:multfriend} compares the friends of multiple-time jnsulting account holders and targets. It is evident from the distribution that account holders have larger number of friends in general with high probability ($p$-value of significance is $<10^{-5}$). 

\section{Discussions and Conclusions}
\subsection{Implications of our work} There are several implications of this study. We report these in the following.\\

\noindent{\textbf{Mitigating the spread of negativity:} Incivility has high impact on the spread of negativity among communities in social media platforms. In prior research works~\cite{cheng2017anyone,willer2009false,cheng2014community}, it has been shown that negative behavior can persist and spread across a community (if not controlled) due to its reinforcing nature. Opinion disagreements can act as early indicators for stopping the spread of such negativities. Our proposed framework can be efficiently used to detect opinion disagreements that might lead to incivilities early in time with minimal features. Since lot of conversations around controversial topics or policy issues, e.g., global warming, vaccination, abortion, immigration etc. on social media generally lead to opinion disagreements, our framework can be tuned to detect the level of toxicity or incivility in the posts so that only the ones which are below a threshold (can be a system parameter) are detected. This will increase the system efficiency of detecting potential incivil incidents. Early detection of these opinion disagreements can act as alerts that can be sent to content moderators (bots or humans) so that they monitor the conflicts and take necessary actions to mitigate the further spread of negativity in the community.\\

\noindent\textbf{Social and policy implications:} Our work also has important implications to law, public policy and the society. Wolfson in context of American concept of free speech and political scenario, has argued in his book~\cite{wolf} that it is almost impossible to separate the good speech or the bad speech/incivility/abusiveness. According to the First Amendment to the United States Constitution, one has to provide adequate opportunities to express differing opinions and engage in public political debate. However, Wolfson also notes that in case of private individuals, these differing opinions impact emotional health of the targeted individual and hence must be prohibited. In contrast, incivility directed toward a group or community has the potential to mobilize a larger number of individuals and can have devastating consequences. These open up the debate for the need to censorship, regulations or Government laws. \\

\noindent\textbf{Design of online platforms to reduce conflicts:}
It might not be always appropriate to attribute the strong sentiment expressed by a target to be representative of his/her internal characteristics; in contrast, it might be because of a very specific and one-time external factor, for instance a ``very bad day'' of the target for which he/she might hold the  named entity responsible eventually causing him/her to post a (one-off) tweet expressing a strong sentiment toward that named entity. Our work sheds light on how we might design discussion platforms that minimize the number of opinion conflicts due to such incidents.

Early detection of opinion conflict might be used by the platform moderators to alter the context of the discussion by possibly selectively hiding strong sentiments toward entities and prioritizing constructive comments. This may directly enhance civility in the platform making account holders less likely to be invective (similar observations in the context of trolling and online harassment has been made in earlier works~\cite{Blackwell:2017,cheng2017anyone}). 

In addition, community norms might be introduced to reduce conflicts. For instance, the moment someone is found to express strong sentiments toward a certain entity, the platform may raise alerts/reminders of ethical standards or cite past moral actions. 

We believe that many of these efforts can 
come directly from the researchers and designers in the CSCW community so as to make the online experience of user more safe and accommodating.

\subsection{Conclusions}
In this paper, we study the behavioral aspects of the targets and account holders and try to understand various factors associated with incivility incidents. We then automate the incivility detection process by developing a deep learning model with character level LSTM and CNNs and incorporating information related to entity-specific opinion conflicts. Our model achieves an accuracy of 93.3\% with an F1-score of 0.82 which significantly outperforms the best performing baseline models achieving 4.9\%, 6.5\% improvements w.r.t accuracy, F1-score respectively. 

\noindent\textit{Insights}: We gained several insights from our analysis and automation. Some of these are -\\
(i) Opinion conflicts among the target and the account holder at early stages can lead to eventual instances of incivility;\\
(ii) Targets are usually more popular with higher reputation compared to account holders; this is opposite to instances of cyberbullying where the perpetrator is usually more reputed and powerful;\\
(iii) Account holders tend to post tweets that are inflicted with negative sentiments/emotions, ``swear'', ``sex'', ``religion'' and ``death'' words;\\
(iv) Targets usually post tweets rich in positive emotions;\\
(v) There are extensive evidences of repeated incivility as well as repeated targeting, sometimes each of these up to 10 times. This could be very alarming for the social media sites and suitable automatic procedures should be built in to contain such cases early in time. Our deep learning model could be a first step toward this enterprise.

\subsection{Future works} There are quite a few interesting directions that can be explored in future. One such direction could be to look into the problem of incivility from perspectives of various \textit{events} and \textit{topics} and study the effect of such topics/events on the incivility dynamics. Another interesting direction could be to study the temporal characteristics of incivility incidents on social media and identify the factors leading to rise in incivility incidents ~\cite{mariabully,rawhide,anabully}. 
Apart from understanding the cause of incivility and thereby detecting it, one can also develop alert-systems that can subsequently take action to mitigate the impact of opinion disagreements leading to incivilities.

\bibliographystyle{ACM-Reference-Format}
\bibliography{reference,reference2}

\end{document}